\documentclass[aps, prd, amsmath, floats,floatfix,  
twocolumn, superscriptaddress, nofootinbib, showpacs]{revtex4}

\usepackage{graphicx} %
\usepackage{color} %
\usepackage{bm} %

\usepackage{amsthm} %
\usepackage{amssymb} %
\usepackage{amsmath} %

\usepackage{braket}
\usepackage{accents} %
\usepackage{xspace} 

\usepackage{url}

\input{Preamble}

\newcommand{\pg}{\ensuremath{\gamma}}
\newcommand{\cg}{\ensuremath{\tilde{\gamma}}}
\newcommand{\ta}{\ensuremath{\tilde{A}}}
\newcommand{\tD}{\ensuremath{\tilde{D}}}

\newcommand{\tg}{\ensuremath{\tilde{\Gamma}}}
\newcommand{\tR}{\ensuremath{\tilde{R}}}
\newcommand{\rp}{\ensuremath{R^\phi}}

\newcommand{\uz}{\mbox{\em \r{u}\hspace{0.3mm}}}

\newcommand{\beq}{\begin{equation}}
\newcommand{\eeq}{\end{equation}}
\newcommand{\bea}{\begin{eqnarray}}
\newcommand{\eea}{\end{eqnarray}}

\definecolor{NoteColor}{rgb}{0,0,.85}

\definecolor{NewColor}{rgb}{0,.55,0}

\makeatletter
\let\protect\relax
{\catcode`\|=\active
  \xdef\InnerProduct{\protect\expandafter\noexpand\csname InnerProduct \endcsname}
  \expandafter\gdef\csname InnerProduct \endcsname#1{%
    \begingroup
    \ifx\SavedDoubleVert\relax
    \let\SavedDoubleVert\|\let\|\IpDoubleVert
    \fi
    \mathcode`\|32768\let|\IPVert
    \left({#1}\right)
    \endgroup
  }
}
\def\IPVert{\@ifnextchar|{\|\@gobble}
     {\egroup\,\mid@vertical\,\bgroup}}
\def\IPDoubleVert{\egroup\,\mid@dblvertical\,\bgroup}
\let\SavedDoubleVert\relax
\def\midvert{\egroup\mid\bgroup}
\def\SetVert{\@ifnextchar|{\|\@gobble}
    {\egroup\;\mid@vertical\;\bgroup}}
\def\SetDoubleVert{\egroup\;\mid@dblvertical\;\bgroup}
\def\mid@vertical{\mskip1mu\vrule\mskip1mu}
\def\mid@dblvertical{\mskip1mu\vrule\mskip2.5mu\vrule\mskip1mu}
\makeatother

\newcommand{\CITA}{\affiliation{Canadian Institute for Theoretical
    Astrophysics, 60 St.~George Street, University of Toronto,
    Toronto, ON M5S 3H8, Canada.}}

\newcommand{\Cornell}{\affiliation{Center for Radiophysics and Space
    Research, Cornell University, Ithaca, New York, 14853, USA.}}
\newcommand{\umd}{\affiliation{Department of Physics, Maryland Center for Fundamental Physics, 
Joint Space Science Institute, Center for Scientific Computation and Mathematical Modeling. University of Maryland, College Park, MD 20742, USA.}}
\newcommand{\mexico}{\affiliation{Instituto de F\'{\i}sica y Matem\'aticas,
Universidad Michoacana de San Nicol\'as de Hidalgo,
Edificio C-3, Ciudad Universitaria, 58040 Morelia, Michoac\'an, M\'exico.}}
\newcommand{\ncsu}{\affiliation{Department of Physics, North Carolina State University,
Raleigh, NC 27695, USA.}}
\newcommand{\lsu}{\affiliation{Center for Computation \&
    Technology, Department of Physics \& Astronomy, Louisiana State University, Baton Rouge, LA 70803, USA.}}
\newcommand{\PI}{\affiliation{Perimeter Institute for Theoretical
    Physics, Waterloo, ON N2L 2Y5, Canada.}}
\newcommand{\guelph}{\affiliation{Department of Physics, University of
    Guelph, Guelph, ON N1G 2W1, Canada.}}
\newcommand{\brownmath}{\affiliation{Division of Applied Mathematics, Brown University, Providence, RI 02912, USA}}
\newcommand{\brownphys}{\affiliation{Department of Physics, Brown University, Providence, RI 02912, USA}}

\sloppypar

\begin{document}


\graphicspath{{Plots/}}

\title{Numerical simulations with a first order BSSN formulation of Einstein's field equations}

\author{J.~David Brown} \ncsu
\author{Peter Diener} \lsu
\author{Scott E.~Field} \umd
\author{Jan S. Hesthaven} \brownmath
\author{Frank Herrmann} \umd
\author{Abdul H.~Mrou\'e} \CITA \Cornell
\author{Olivier Sarbach} \mexico 
\author{Erik Schnetter} \PI \guelph \lsu
\author{Manuel Tiglio} \umd 
\author{Michael Wagman} \brownphys

\date{\today}

\begin{abstract}
  We present a new fully first order strongly hyperbolic representation of the
  BSSN formulation of Einstein's equations with optional constraint damping terms. We 
  describe the characteristic fields of the system, discuss its hyperbolicity properties, and present two numerical implementations and simulations: one using finite differences, adaptive mesh refinement and in  particular binary black holes, and another one using the discontinuous Galerkin method in spherical symmetry. 
   The results of this paper  constitute a  first step in an effort to combine the robustness of BSSN evolutions with very high accuracy numerical techniques, such as spectral collocation multi-domain or discontinuous Galerkin methods. 
\end{abstract}


\maketitle

\section{Introduction}\label{Sec:Introduction}

Complete, long term numerical simulations of the inspiral, merger and ringdown of two black holes became possible a few years ago~\cite{2005PhRvL..95l1101P,Campanelli:2005dd,Baker:2005vv} and are now carried out by numerous groups; see \cite{Pretorius:2007nq,Centrella:2010mx,Marronetti:2011pv} for recent reviews on the topic. One of the motivations for studying the dynamics of these inspiraling compact binaries is due to the fact that they are among the most promising sources of gravitational waves for the upcoming advanced network of earth based laser interferometric detectors \cite{Abadie:2010cfa}. Moreover, with (and only with) modeling of enough accuracy, these  detectors should be able to extract from the waves physical data about these sources such as the component masses and spins. 

Until a few years ago, such simulations were plagued by short-term instabilities.  With full, long and stable simulations now being carried out systematically, development efforts have focused on efficiency and accuracy, better boundary conditions (see \cite{Reula:2010yt} for a review), and wave extraction methods (see, for example, \cite{Boyle:2009vi,Pollney:2009ut,Reisswig:2009rx,Babiuc:2010ze} and references therein), all of which are especially important for many-orbit evolutions, such as those needed to make comparisons with post-Newtonian models, and calibration or fitting of semi-analytical or phenomenological models \cite{Blanchet:LRR,Baker:2006ha,Buonanno:2007pf,Damour:2007yf,Ajith:2007kx,Damour:2007nc,Damour:2007vq,Boyle:2007ft,Berti:2007fi,Hannam:2007ik,Hannam:2007wf,Gopakumar:2007vh,Damour:2008gu,Boyle:2008ge,Hannam:2009hh,Santamaria:2010yb,Buonanno:2009qa,Pan:2009wj,Pan:2010hz,Pan:2011gk,Barausse:2011ys}. We consider here only solutions of the vacuum Einstein equations, and
intentionally ignore the much larger and astrophysically probably even
more interesting case where matter, radiation, or electromagnetic
fields are present.

Most, if not all, numerical simulations of binary black holes (BBH) currently use one of two formulations of the Einstein equations. One of them is the  generalized harmonic system, which has been successfully implemented in BBH simulations using finite difference adaptive mesh refinement (AMR) \cite{Pretorius:2007nq}, pseudospectral collocation codes (see, for example, \cite{Scheel:2006gg,Scheel:2008rj,Chu:2009md,Szilagyi:2009qz,Lovelace:2011nu} and references therein), and multi-domain finite differences \cite{Pazos:2009vb}, in a first order in time and second order in space formulation~\cite{Pretorius05} in the AMR case, and a fully first order reduction~\cite{Lindblom:2005qh} otherwise. In either case the key ingredient is a constraint damping mechanism \cite{Gundlach:2005eh}, originally proposed in \cite{Brodbeck:1998az} (in that reference referred to as $\lambda$-systems because they were first introduced as Lagrange multipliers).  The other one is the Baumgarte--Shapiro--Shibata--Nakamura (BSSN) system~\cite{shibata95,baumgarte99}, which has been implemented by many groups using finite difference codes in a first order in time, second order in space form (see \cite{Centrella:2010mx} for a review)-- we refer to this as simply {\it BSSN} or {\it second-order BSSN} (as opposed to our fully first order reduction, to which we will refer as {\it FOBSSN}).  Some variant of the ``standard gauge'' conditions for the BSSN formulation, consisting of 1+log slicing or generalizations thereof and the so called Gamma--driver shift~\cite{Alcubierre02a} condition are, more often than not, used. The hyperbolicity of BSSN with a generalization of these gauge conditions was studied in Refs.~\cite{Beyer:2004sv,Gundlach:2006tw}.

While the Einstein equations are fundamentally a second order system, many advanced numerical techniques for hyperbolic systems -- such as multi-domain high order finite difference, spectral collocation, and discontinuous Galerkin methods -- are well developed for first order hyperbolic systems. At the same time, the standard second order in space BSSN system with the standard gauge conditions has shown remarkably robust properties in a variety of compact binary configurations. One wonders, then, if it is possible to combine some of the numerical techniques that are often used for very high accuracy simulations of hyperbolic differential equations with the BSSN system. 

One recent approach has been to adapt advanced techniques for fully first order systems to second order in space ones \cite{Tichy:2006qn,Tichy:2009zr,Field:2010mn,Taylor:2010ki,cattoenThesis2009}. Perhaps paradoxically, it appears to be more difficult to guarantee stability for naturally second order systems than for first order reductions of them, though progress is being made on this front (see, for example, \cite{Taylor:2010ki,Cecere:2011aa,it:2011-008}). Another approach is to rewrite the Einstein equations as a fully first order hyperbolic system. In this paper we explore the latter and we refer to our first order reduction of BSSN as FOBSSN\@.

This paper is organized as follows. Section~\ref{sec:BSSN} reviews the BSSN system in covariant form. The first order reduction is carried out in Section~\ref{sec:FOBSSN}, where we also show that FOBSSN is strongly hyperbolic under suitable conditions on the gauge parameters, and discuss the propagation of the constraints. Section~\ref{sec:Results} summarizes some of our results from numerical simulations of binary black holes using FOBSSN and adaptive-mesh refinement and finite differences,  and a multi-domain nodal discontinuous Galerkin scheme in spherical symmetry. When appropriate, we compare results from our FOBSSN simulations to simulations using BSSN in its standard form. A preliminary look at turduckening \cite{Etienne:2007hr,Brown:2007pg,Brown:2008sb} for a polynomial/spectral Galerkin method is presented in the context of the spherically reduced FOBSSN system. Appendices collect further details on the covariant BSSN system as well as expressions for the fundamental fields in terms of characteristic variables.

\section{Review of the BSSN System} \label{sec:BSSN}
We briefly review the second order form of 
BSSN~\cite{shibata95,baumgarte99} with moving puncture gauge conditions. Here we follow the approach of Ref.~\cite{Brown:2009dd}, which is spatially-covariant (but not fully space-time covariant). The spatial metric and extrinsic curvature are denoted $\gamma_{ij}$ and $K_{ij}$, respectively. These are replaced by the BSSN variables
\begin{subequations}\label{BSSNvariabledef}
\begin{eqnarray}
  \phi &=& \frac{1}{12} \ln (\gamma/\overline\gamma)  \,, \\
  K &=& \pg^{ij} K_{ij}              \,, \\
  \cg_{ij} &=& e^{- 4 \phi} \pg_{ij}  \,, \\
  \tilde A_{ij} &=& e^{- 4 \phi} \left( K_{ij} - \frac{1}{3}\gamma_{ij} K\right) \,,
\end{eqnarray}
\end{subequations}
where $\gamma$ is the determinant of $\gamma_{ij}$ and $\overline\gamma$ is a fiducial scalar density of weight $2$ which remains to be specified. Also, $K = \gamma^{ij}K_{ij}$ is the trace of the extrinsic curvature. The variable $\tilde\gamma_{ij}$ is the conformal metric, $\phi$ is the conformal factor, and $\tilde A_{ij}$ is the conformally rescaled, trace--free part of the extrinsic curvature. 
(In addition to this ``$\phi$'' variant of BSSN, there exist also the
$W$ and $\chi$ variants where the variable $\phi$ is replaced by
$W=(\gamma/\overline\gamma)^{-1/6}$ or
$\chi=(\gamma/\overline\gamma)^{-1/3}$, respectively.)
These variables are restricted by the algebraic conditions
\begin{eqnarray}
  \label{eq:alg-constr}
  \tilde\gamma = \overline\gamma \ , \qquad \tilde\gamma^{ij}\tilde A_{ij} = 0 \ ,
\end{eqnarray}
where $\tilde\gamma$ is the determinant of $\tilde\gamma_{ij}$.\footnote{Note that we use both $\tilde\gamma$ and $\overline\gamma$ in our notation.} The BSSN system also includes the ``conformal connection vector'' 
\begin{equation} \label{eq:Gamma}
    \tilde\Lambda^i = \tilde\gamma^{jk} \Delta\tilde\Gamma^i{}_{jk} \ ,
\end{equation}
as independent variables, and we have defined $\Delta\tilde\Gamma^i{}_{jk} \equiv \tilde\Gamma^i{}_{jk} - \overline\Gamma^i{}_{jk}$. Here, $\tilde\Gamma^i{}_{jk}$ are the Christoffel symbols built from the conformal metric and $\overline\Gamma^i{}_{jk}$ is a fiducial connection. In this covariant language the BSSN variables are tensors with no density weights. In particular, $\phi$ is a scalar and $\tilde{\Lambda}^i$ is a contravariant vector.

It is often convenient to consider the fiducial connection $\overline\Gamma^i{}_{jk}$ to be constructed from a ``fiducial metric'' $\overline\gamma_{ij}$ whose determinant is $\overline\gamma$. We stress that the fiducial fields are not dynamical variables. They are freely chosen functions, required by covariance. Throughout the main body of the paper we assume that the fiducial connection is built from a flat, time independent metric $\overline\gamma_{ij}$ whose determinant is $\overline\gamma$. If the coordinates are interpreted as Cartesian, then $\overline\gamma_{ij} = {\rm diag}(1,1,1)$. In this case $\overline\gamma = 1$ and $\overline\Gamma^i{}_{jk} = 0$. The vector $\tilde\Lambda^i$ then reduces to the conformal connection functions $\tilde\Gamma^i \equiv \tilde\Gamma^i{}_{jk}\tilde\gamma^{jk}$ and Eqs.~\eqref{eq:evolBSSN} below reduce to the usual second order BSSN system.

The evolution equations for the BSSN variables are 
\begin{subequations} \label{eq:evolBSSN}
\begin{eqnarray}
\partial_\perp \phi & = & \frac{1}{6} \overline D_i \beta^i
	-\frac{1}{6}\alpha K \ ,\label{eq:evolphi}\\
\partial_\perp \tilde{\gamma}_{ij} & = & -\frac{2}{3}\tilde\gamma_{ij} \overline D_k\beta^k 
	-2\alpha\tilde{A}_{ij} \ ,\label{eq:evolg}\\
\partial_\perp K & = &
	\alpha\left(\tilde{A}_{ij}\tilde{A}^{ij}+\frac{1}{3}K^2
	\right)-\gamma^{ij}D_iD_j\alpha  \label{eq:evolK} \ ,\\
\partial_\perp \ta_{ij} & = & -\frac{2}{3}\tilde A_{ij} \overline D_k\beta^k +
	\alpha\left(K\ta_{ij}-2\ta_{ik} 
   	\ta^k{}_j\right)\nonumber\\ 
   	&&+e^{-4\phi}\left[\alpha R_{ij} -D_iD_j\alpha\right]^{\text{TF}} \ ,
   	\label{eq:evolA} \\
\partial_\perp\tilde\Lambda^i & = &
		\tilde\gamma^{k\ell}\overline D_k \overline D_\ell \beta^i 
		+ \frac{2}{3}\tilde\gamma^{jk}\Delta\tilde\Gamma^i{}_{jk} \overline D_\ell\beta^\ell \nonumber\\ &&
		+ \frac{1}{3}\tilde D^i(\overline D_k\beta^k) 
		- 2\tilde A^{ik} \overline D_k\alpha
		+ 2\alpha \tilde A^{k\ell}\Delta\tilde\Gamma^i{}_{k\ell}  \nonumber \\ && 
		+ 12\alpha \tilde A^{ik}\overline D_k\phi
		- \frac{4}{3} \alpha \tilde D^i  K  \ , \label{eq:evolGam}
\end{eqnarray}
\end{subequations}
where $\alpha$ is the lapse function and $\beta^i$ is the shift vector. Also, the time derivative operator is defined by $\partial_\perp \equiv \partial_t - {\cal L}_\beta$, where ${\cal L}_\beta$ is the Lie derivative with respect to the shift. Next, $D_i$, $\tilde D_i$ and $\overline D_i$ are the covariant derivatives built from the physical metric, conformal metric, and fiducial metric, respectively, and $D_i D_j\alpha = \overline D_i\overline D_j\alpha - \Delta\Gamma^k{}_{ij}\overline D_k\alpha$. Finally, in Eq.~(\ref{eq:evolA}),  TF denotes the trace-free part of the expression in brackets. 

The Ricci tensor can be written as a sum of two terms, 
\begin{equation}\label{Ricciconformalsplit}
   R_{ij} = \tR_{ij} + \rp_{ij} \ .
\end{equation}
The Ricci tensor for the conformal metric is  
\begin{eqnarray}\label{conformalRicci}
   \tR_{ij}  & = & -\frac{1}{2} \tilde\gamma^{k\ell} \overline D_k \overline D_\ell \tilde\gamma_{ij} 
      + \tilde\gamma_{k(i} \overline D_{j)} \tilde\Lambda^k 
		+\tilde\gamma^{lm} \Delta\tilde\Gamma^k{}_{lm} \Delta\tilde\Gamma_{(ij)k}
		\nonumber\\ &&
		+ \tilde\gamma^{k\ell} [ 2\Delta\tilde\Gamma^m{}_{k(i} \Delta\tilde\Gamma_{j)m\ell} 
			+ \Delta\tilde\Gamma^m{}_{ik} \Delta\tilde\Gamma_{mj\ell} ]  \ ,
\end{eqnarray}
and the term  $R^\phi_{ij}$ is defined by 
\begin{eqnarray}\label{Ricciphiterms}
   \rp_{ij} &=& -2\tD_i\tD_j\phi-2 \cg_{ij}
    \tD^k\tD_k \phi \nonumber\\
   &&+4\tD_i\phi\tD_j\phi-4\cg_{ij}\tD^l\phi \tD_l\phi \ .
\end{eqnarray}
All tensors with a tilde have their indices raised and lowered with the conformal metric$\cg_{ij}$. Details of the derivation of the equations of motion (\ref{eq:evolBSSN}) are contained in Appendix \ref{app:BSSN}.

In addition to the algebraic constraints, solutions to the second order BSSN system must satisfy a set of differential constraints stemming from the $3+1$ decomposition (Hamiltonian and momentum constraints) and definition of the conformal connection vector. Expressed in terms of the evolved variables, these are given by
\begin{subequations}\label{secondorderconstraints}
\begin{eqnarray}
{\mathcal H} &=& e^{-4\phi}\left(\tilde{R}-8\tilde{D}^i\tilde{D}_i\phi
	-8\tilde{D}^i\phi\tilde{D}_i\phi\right) 
   \nonumber\\ && 
   +\frac{2}{3}K^2-\tilde{A}_{ij}\tilde{A}^{ij} =0 \ , \\
\tilde{{\mathcal M}}^i &=& \tilde{D}_j  \tilde{A}^{ij}
	+6\tilde{A}^{ij}\partial_j\phi -\frac{2}{3} \tilde D^i K  = 0 \ , \\
{\mathcal G}^i &=& \tilde \Lambda^i  - \tilde \gamma^{jk} \Delta\tilde\Gamma^i_{jk} = 0 \ ,
\end{eqnarray}
\end{subequations}
where $\tilde{R}=\tilde{\gamma}^{ij}\tilde{R}_{ij}$. Using the Bianchi identities and the BSSN equations (\ref{eq:evolBSSN}), one can derive a closed homogeneous evolution system for the constraint fields ${\mathcal H}$, $\tilde{{\mathcal M}}^i$ and ${\mathcal G}^i$. This constraint evolution system can be written in first order symmetric hyperbolic form~\cite{Beyer:2004sv,Gundlach:2004jp,Brown:2008sb,Nunez:2009wn}. Therefore, if the initial data satisfies the constraints, then the constraints will be preserved for all times as long as suitable boundary conditions are provided. Constraint--preserving boundary conditions for the BSSN system have been discussed in Refs.~\cite{Gundlach:2004jp,Nunez:2009wn}. 

Black hole evolutions with the second--order BSSN system are typically carried out using the moving puncture gauge conditions consisting of 1+log slicing and Gamma--driver shift. In this paper we consider the general Bona-Mass\'o slicing condition~\cite{Bona94b}, written in the form~\cite{Beyer:2004sv}
\begin{equation}
\partial_\perp \alpha = -\alpha^2 f(\alpha, \phi) K + S_\alpha \ ,
\label{eq:bm-slicing}
\end{equation} 
where $f(\alpha,\phi)$ is an arbitrary positive function of the lapse $\alpha$ and the conformal factor $\phi$. The source term $S_\alpha$ is a function of the spatial coordinates. For $1+\log$ slicing, $f(\alpha, \phi) = 2/\alpha$ and $S_\alpha=0$. 

The shift condition considered in this paper is a generalization of the Gamma--driver shift, written as~\cite{Beyer:2004sv}
\begin{subequations}\label{gammadrivershifteqns}
\begin{eqnarray}
	\overline\partial_0\beta^i & = & \alpha^2 G(\alpha,\phi) B^i + S_\beta^i \ ,\label{eq:evolbeta}\\
	\overline\partial_0 B^i & = & e^{-4\phi} H(\alpha,\phi) \overline\partial_0 \tilde\Lambda^i - \eta B^i + S_B^i \label{eq:evolB} \ .
\end{eqnarray}
\end{subequations}
Here, $B^i$ is an auxiliary field and the time derivative operator is defined by $\overline\partial_0 \equiv \partial_t - \beta^j\overline D_j $. The functions $G$ and $H$ depend on the lapse $\alpha$ and conformal factor $\phi$. The source terms $S_\beta^i$ and $S_B^i$ are functions of the spatial coordinates. The standard choices for the Gamma--driver shift condition are $G(\alpha,\phi)=3/(4 \alpha^2)$, $H(\alpha,\phi)=e^{4\phi}$, $S_\beta^i = S_B^i = 0$, and $\eta = 3/(4M)$, where $M$ is the ADM mass of the system or another relevant mass scale. (If different regions of the domain have different mass scales, e.g.\ for binary black hole systems with a large mass ratio, then $\eta$ may vary in space \cite{Muller:2009jx, Schnetter:2010cz}.)

\section{First-Order BSSN}\label{sec:FOBSSN}
\subsection{Evolution System}
The BSSN system as described above contains second order derivatives in space acting on the variables $\alpha$, $\beta^i$, $\phi$ and $\tilde\gamma_{ij}$. To write the system in fully first order form we introduce the new variables
\begin{subequations}
\begin{eqnarray}
& \alpha_i = \overline D_i\alpha \ , \\
& \beta_i{}^j = \overline D_i\beta^j \ ,\\
& \phi_i = \overline D_i\phi \ , \\
& \tilde\gamma_{kij} = \overline D_k \cg_{ij} \ .
\end{eqnarray}
\end{subequations}
These definitions yield the associated constraints
\begin{subequations}
\begin{eqnarray}
& {\cal A}_i \equiv \alpha_i - \overline D_i \alpha = 0 \ , \label{Aconstraint}\\
& {\cal B}_i{}^j  \equiv \beta_i{}^j - \overline D_i \beta^j = 0 \ ,\label{Bconstraint}\\
& {\cal C}_i \equiv \phi_i - \overline D_i \phi = 0 \ ,\label{Cconstraint}\\
& {\cal D}_{kij} \equiv \tilde\gamma_{kij} - \overline D_k \cg_{ij} = 0 \ .\label{Dconstraint}
\end{eqnarray}
\end{subequations}
Observe that the derivative $\overline D_k$ applied to the algebraic constraint $\tilde\gamma = \overline\gamma$ 
yields the condition 
\begin{equation}\label{tracegammakij}
        \tilde\gamma^{ij} \tilde\gamma_{kij} = 0 \ .
\end{equation}
This is a new algebraic constraint that the first order BSSN variables must satisfy, along with the algebraic constraints (\ref{eq:alg-constr}) inherited from second order BSSN. 
  
The evolution equations for the new variables are obtained by computing their time derivatives using the second--order BSSN equations (\ref{eq:evolphi}), (\ref{eq:evolg}) and gauge conditions (\ref{eq:bm-slicing}), (\ref{eq:evolbeta}). In carrying out these calculations we continue to assume that the fiducial metric is flat and time--independent. The complete system of first order equations, the FOBSSN system, can be conveniently split into gauge and non--gauge sectors.  The gauge sector is
\begin{widetext}
\begin{subequations}\label{gaugesector}
\begin{eqnarray}
\overline\partial_0 \alpha &=& -\alpha^2 f K  + S_\alpha \ ,
\label{eq:alpha-evol}
\\
\overline\partial_0 \alpha_i &=& -\alpha^2 f \partial_i K 
 - 2f\alpha\alpha_i K 
 - \alpha^2\left[ f_\alpha\alpha_i +  f_\phi \phi_i \right] K 
+ \beta_i{}^j\alpha_j + \overline D_i S_\alpha - \kappa^\alpha {\cal A}_i \ ,
\label{eq:alpha_i-evol}
\\
\overline\partial_0 \beta^j &=& \alpha^2 G B^j + S_\beta^j \ ,
\label{eq:beta-evol}
\\
\overline\partial_0 B^j &=& e^{-4\phi} H
 \overline\partial_0 \tilde\Lambda^j - \eta B^j + S_B^j\ ,
\label{eq:B-evol}
\\
\overline\partial_0 \beta_i{}^j &=& \alpha^2 G\overline D_i B^j
 + 2G\alpha\alpha_i B^j
 + \alpha^2\left[ G_\alpha\alpha_i + G_\phi \phi_i \right] B^j
+ \beta_i{}^k\beta_k{}^j + \overline D_i S_\beta^j - \kappa^\beta {\cal B}_i{}^j \ ,
\label{eq:betai-evol}
\\
\overline\partial_0 K &=& -e^{-4\phi}\cg^{ij}
 \left[ \tilde D_i \alpha_j 
 + 2 \phi_i\alpha_j \right]
 + \alpha\left( \ta^{ij}\ta_{ij} + \frac{1}{3}\, K^2 \right) \ .\label{eq:K-evol}
\end{eqnarray}
\end{subequations}
\end{widetext}
Here, subscripts $\alpha$ and $\phi$ on the functions $f$ and $G$ denote partial derivatives. Note that terms proportional to the constraints ${\cal A}_i$ and ${\cal B}_i{}^j$ have been added to the right--hand sides of the evolution equations for $\alpha_i$ and $\beta_i{}^j$. The corresponding proportionality constants are $\kappa^\alpha$ and $\kappa^\beta$. These terms can be used as a damping mechanism for any numerical violation of the constraints ${\cal A}_i=0$ and ${\cal B}_i{}^j=0$.

The remaining evolution equations, which comprise the non--gauge sector, are 
\begin{widetext}
\begin{subequations}\label{nongaugesector}
\begin{eqnarray}
\overline\partial_0 \phi &=& -\frac{\alpha}{6}\, K + \frac{1}{6}\beta_k{}^k \ ,\label{eq:phi-evol}
\\
\overline\partial_0 \phi_i &=& -\frac{1}{6}\alpha \overline D_i K 
+ \frac{1}{6}\overline D_i \beta_k{}^k
 - \frac{1}{6}\alpha_i K + \beta_i{}^j \phi_j - \kappa^\phi {\cal C}_i \ ,
\\
\overline\partial_0 \cg_{ij} &=& -2\alpha\ta_{ij}
 + 2\tilde\gamma_{k(i}\beta_{j)}{}^k - \frac{2}{3} \tilde\gamma_{ij} \beta_k{}^k \ ,\label{eq:gammatilde-evol}
\\
\overline\partial_0 \ta_{ij} &=& e^{-4\phi}\left[
 \alpha\tR_{ij} - 2\alpha \overline D_{(i}\phi_{j)} + 4\alpha \phi_i\phi_j 
 - \overline D_{(i}\alpha_{j)} + \Delta\tilde\Gamma^k{}_{ij}(2\alpha\phi_k + \alpha_k)
 + 4\alpha_{(i}\phi_{j)} \right]^{TF} 
\nonumber\\
 & & + \alpha K\ta_{ij} - 2\alpha\ta_{ik}\ta^k_{\; j}
  + 2\ta_{k(i}\beta_{j)}{}^k - \frac{2}{3}\ta_{ij}\beta_k{}^k 
\ ,\label{eq:Atilde-evol}
\\
\overline\partial_0 \tilde\gamma_{kij} &=& -2\alpha\overline D_k\ta_{ij}
 + 2 (\overline D_k \beta_{(i}{}^\ell)\cg_{j)\ell} - \frac{2}{3}\tilde\gamma_{ij} \overline D_k \beta_\ell{}^\ell
\nonumber\\
& & - 2\alpha_k\ta_{ij} + \beta_k{}^\ell\tilde\gamma_{\ell ij}
 + 2\tilde\gamma_{k\ell (i}\beta_{j)}{}^\ell - \frac{2}{3}\, \tilde\gamma_{kij}\beta_\ell{}^\ell
 - \kappa^\gamma  {\cal D}_{kij} \ ,\label{eq:dkijtilde-evol}
\\
\overline\partial_0 \tilde\Lambda^i &=& \cg^{k\ell}\overline D_k \beta_\ell{}^i
 + \frac{1}{3}\tilde D^i \beta_k{}^k 
 + \sigma\cg^{ij}
   \left( \overline D_j \beta_k{}^k - \overline D_k \beta_j{}^k \right)
 - \frac{4}{3}\alpha \tilde D^i K
\nonumber\\
& & - \Delta\tg^\ell{}_{jk}\cg^{jk}\beta_\ell{}^i
 + \frac{2}{3}\Delta\tg^i{}_{jk}\cg^{jk}\beta_\ell{}^\ell
 - 2\ta^{ij}\alpha_j
 + 2\alpha\left( \Delta\tg^i{}_{k\ell}\ta^{k\ell}
   + 6\ta^{ij} \phi_j \right) \ .
\label{eq:gamma-evol}
\end{eqnarray}
\end{subequations}
Here, we have defined 
\begin{eqnarray*}
\Delta\tilde{\Gamma}^i{}_{k\ell} &=& \frac{1}{2}\tilde{\gamma}^{ij}\left(
 \tilde{\gamma}_{k\ell j} + \tilde{\gamma}_{\ell kj} - \tilde{\gamma}_{jk\ell} \right) \ , \nonumber\\
    \tR_{ij}  & = & -\frac{1}{2} \tilde\gamma^{k\ell} \overline D_k \tilde\gamma_{\ell ij} 
      + \tilde\gamma_{k(i} \overline D_{j)} \tilde\Lambda^k 
		+ \tilde\gamma^{\ell m}\Delta\tilde\Gamma^k_{\ell m} \Delta\tilde\Gamma_{(ij)k}
		+ \tilde\gamma^{k\ell} [ 2\Delta\tilde\Gamma^m{}_{k(i} \Delta\tilde\Gamma_{j)m\ell} 
			+ \Delta\tilde\Gamma^m{}_{ik} \Delta\tilde\Gamma_{mj\ell} ] 
		  \ ,
\nonumber\\
\end{eqnarray*}
\end{widetext}
which follow from the definition of the Christoffel symbols and the identity (\ref{conformalRicci}) for the Ricci tensor. In the evolution equations for $\phi_i$ and $\tilde \gamma_{kij}$, the constraints ${\cal C}_i$ and ${\cal D}_{kij}$ are subtracted with constants $\kappa^\phi$ and $\kappa^\gamma$. These terms are included as damping mechanisms for these constraints, in analogy with $\kappa^{\alpha}$ and $\kappa^{\beta}$. The term proportional to the constant $\sigma$ in the evolution equation for $\tilde{\Lambda}^i$, Eq.~\eqref{eq:gamma-evol},  is equal to the constraint $2\tilde{\gamma}^{ij}\overline D_{[j} {\cal B}_{k]}{}^k = 0$. This term is needed to make the evolution system strongly hyperbolic, as discussed below.

\subsection{Constraint Propagation}
\label{SubSec:CP}

The FOBSSN system is subject to the algebraic constraints $\tilde\gamma  - \overline\gamma = 0$, $\tilde{A}_i^i = \tilde\gamma^{ij}\tilde A_{ij} = 0$, and $\tilde{\gamma}_{ki}{}^i = \tilde\gamma^{ij} \tilde \gamma_{kij} = 0$. As discussed in the next section, our numerical codes enforce some but not all of these constraints. If the algebraic constraints are not enforced, but free to evolve, the first order BSSN evolution equations (\ref{eq:gammatilde-evol}), (\ref{eq:Atilde-evol}) and (\ref{eq:dkijtilde-evol}) imply
\begin{subequations}\label{algebraicconstraintsprop}
\begin{eqnarray}
\overline{\partial}_0 \ln (\tilde{\gamma}/\overline\gamma) &=& -2\alpha\tilde{A}_i^i \ ,
\label{Eq:detgamma}\\
\overline{\partial}_0 \tilde{A}_i^i &=& \alpha K \tilde{A}_i^i \ ,
\label{Eq:trA}\\
\overline{\partial}_0 \tilde{\gamma}_{ki}{}^i &=& -2\alpha\overline D_k \tilde{A}_i^i
- 2\alpha_k \tilde{A}_i^i
+ \beta_k{}^l \tilde{\gamma}_{\ell i}{}^i  \nonumber \\ &&
+ 2\alpha\tilde{A}^{ij}{\cal D}_{kij} - \kappa^\gamma \tilde\gamma^{ij}   {\cal D}_{kij} \ .
\label{Eq:trgamma}
\end{eqnarray}
\end{subequations}
It follows from Eq.~(\ref{Eq:trA}) that $\tilde{A}_i^i$ is zero along an integral curve $c$ of $\overline{\partial}_0 = \partial_t - \beta^j\overline{D}_j$ if it is zero at some point on this curve. Therefore, if all integral curves $c$ intersect the initial surface, it is sufficient to require $\tilde{A}_i^i = 0$ on this surface in order to guarantee that the algebraic constraint $\tilde{A}_i^i = 0$ holds at every time and everywhere on the computational domain. On the other hand, if $c$ intersects the boundary surface, which happens if the shift vector is pointing outwards at the boundary, then $\tilde{A}_i^i = 0$ needs to be enforced as a boundary condition in order to guarantee the satisfaction of the algebraic constraint $\tilde{A}_i^i = 0$. If this constraint holds, it follows by a similar argument from Eq.~(\ref{Eq:detgamma}) that the determinant constraint $\tilde\gamma = \overline\gamma$ holds if it is satisfied initially and suitable boundary conditions are specified in case $\beta^k$ is outward pointing at the boundary. Eqs.~(\ref{Eq:detgamma},\ref{Eq:trA}) also show that is is consistent to enforce the algebraic constraints $\tilde\gamma = \overline\gamma$ and $\tilde{A}_i^i = 0$ throughout evolution, as is the case for the second order BSSN system.

On the other hand, it does not follow immediately from Eq.~(\ref{Eq:trgamma}) and suitable initial and boundary conditions that the trace constraint $\tilde{\gamma}_{ki}{}^i = 0$ holds, unless the term $\tilde{A}^{ij}{\cal D}_{kij}$ is zero.Notice that $\tilde{\gamma}^{ij}{\cal D}_{kij} = \tilde{\gamma}_{ki}{}^i - \overline D_k\ln (\tilde{\gamma}/\overline\gamma)$ so this term can be expressed in terms of the algebraic constraints. This means that the propagation of the algebraic trace constraint $\tilde{\gamma}_{ki}{}^i = 0$ is coupled to those of the constraints ${\cal A}_i = 0$, ${\cal B}_i{}^j = 0$, ${\cal C}_i = 0$ and ${\cal D}_{kij} = 0$, and one cannot consistently enforce $\tilde{\gamma}_{ki}{}^i = 0$ along with the other algebraic constraints unless $\tilde{A}^{ij}{\cal D}_{kij} = 0$.

Alternatively, it is possible to decouple the algebraic constraints from the remaining ones by adding the term
\begin{equation}
-\frac{2\alpha}{3}\tilde{\gamma}_{ij}\tilde{A}^{lm}{\cal D}_{klm}
 + \frac{\kappa^\gamma}{3}\tilde{\gamma}_{ij}\tilde{\gamma}^{lm}{\cal D}_{klm}
\label{Eq:dkijtildeMod}
\end{equation}
to the right-hand side of Eq.~(\ref{eq:dkijtilde-evol}), which has the same effect as the replacements
\begin{eqnarray}
\overline D_k\tilde{A}_{ij} &\mapsto& \overline D_k\tilde{A}_{ij} 
 - \frac{1}{3}\tilde{\gamma}_{ij}\tilde{\gamma}^{lm}\overline D_k\tilde{A}_{lm}
\nonumber\\
 &+& \frac{1}{3}\tilde{\gamma}_{ij}\left[ \overline D_k(\tilde{A}_m^m) 
 + \tilde{A}^{km}\tilde{\gamma}_{klm} \right]
\nonumber
\end{eqnarray}
and
\begin{displaymath}
{\cal D}_{kij} \mapsto {\cal D}_{kij} 
 - \frac{1}{3}\tilde{\gamma}_{ij}\tilde{\gamma}^{lm}{\cal D}_{klm}
\vspace{7 pt}
\end{displaymath}
in that equation. With this, the last two terms on the right-hand side of Eq.~(\ref{Eq:trgamma}) drop, and one obtains a closed, homogeneous evolution system for the algebraic constraints. Therefore, it is consistent to set the algebraic constraints to zero even if ${\cal D}_{kij}\neq 0$.

We now consider the constraints ${\cal A}_i = 0$, ${\cal B}_i{}^j = 0$, ${\cal C}_i = 0$ and ${\cal D}_{kij} = 0$  that were introduced in the reduction of BSSN to first--order. The evolution equations imply that the constraint fields ${\cal A}_i$, ${\cal B}_i{}^j$, ${\cal C}_i$ and ${\cal D}_{kij}$ satisfy the following linear, homogeneous system of equations: 
\begin{widetext}
\begin{subequations}
\begin{eqnarray}
\overline{\partial}_0 {\cal A}_i &=& -2\alpha f K  {\cal A}_i
 - \alpha^2 K \left[ 
      f_\alpha {\cal A}_i + f_\phi {\cal C}_i \right]
 + (\overline D_i\beta^j) {\cal A}_j + \alpha_j{\cal B}_i{}^j
 - \kappa^\alpha {\cal A}_i \ ,
\label{eq:constrA-evol}
\\
\overline{\partial}_0 {\cal B}_i{}^j &=& 2\alpha G B^j {\cal A}_i
 +\alpha^2 B^j\left[ 
     G_\alpha {\cal A}_i + G_\phi {\cal C}_i \right]
 + (\overline D_i\beta^k) {\cal B}_k{}^j + \beta_k{}^j{\cal B}_i{}^k
 - \kappa^\beta {\cal B}_i{}^j \ ,
\label{eq:constrB-evol}
\\
\overline{\partial}_0 {\cal C}_i &=& -\frac{K}{6} {\cal A}_i 
 + (\overline D_i\beta^j) {\cal C}_j + \phi_j {\cal B}_i{}^j - \kappa^\phi {\cal C}_i \ ,
\label{eq:constrC-evol}
\\
\overline{\partial}_0 {\cal D}_{kij} &=& -2\tilde{A}_{ij} {\cal A}_k
 + (\overline D_k\beta^\ell) {\cal D}_{\ell ij} + \tilde{\gamma}_{\ell ij} {\cal B}_k{}^\ell
 + 2{\cal D}_{k\ell(i}\beta_{j)}{}^\ell  - \frac{2}{3}\, {\cal D}_{kij}\beta_\ell{}^\ell
 - \kappa^\gamma {\cal D}_{kij} \ .
\label{eq:constrD-evol}
\end{eqnarray}
\end{subequations}
\end{widetext}
(The term in Eq.~(\ref{Eq:dkijtildeMod}) has to be added to the right-hand side of the last equation in case Eq.~(\ref{eq:dkijtilde-evol}) is modified in the way described above.)

Here, we have set the source terms $S_\alpha$, $S_\beta^i$, and $S_B^i$ to zero for simplicity but without loss of generality with respect to the main conclusions. If the shift is not outward pointing at the boundaries, and the initial data is chosen such that ${\cal A}_i = 0$, ${\cal B}_i{}^j = 0$, ${\cal C}_i = 0$ and ${\cal D}_{kij} = 0$, then these results show that a solution to the first--order BSSN evolution equations will automatically satisfy these constraints for all times. It follows that such a solution will also satisfy the original second order BSSN system. If the shift is outward pointing at a boundary, additional boundary conditions need to be specified in order to ensure that these constraints propagate, see the discussion below Eq.~(\ref{algebraicconstraintsprop}).

On the other hand, numerical errors can trigger small violations of the constraints and these violations might grow in time. We can use the parameters $\kappa$ to help insure that the constraints are damped. As we show in the next subsection, $\kappa^\phi = 0$ is required for strong hyperbolicity, so let us consider $\kappa^\phi = 0$ here as well. Now observe that with the standard gauge conditions the functions $f$ and $G$ are independent of $\phi$. In this case the first two equations, Eqs.~\eqref{eq:constrA-evol}--\eqref{eq:constrB-evol}, decouple from the last two. With $\kappa^\alpha$ and $\kappa^\beta$ sufficiently large, ${\cal A}_i$ and ${\cal B}_i{}^j$ should be damped. Next, observe that the fourth equation, Eq.~\eqref{eq:constrD-evol}, is independent  of ${\cal C}_i$. Assuming ${\cal A}_i$ and ${\cal B}_i{}^j$ are damped, the constraint ${\cal D}_{ijk}$ should remain damped for sufficiently large constant $\kappa^\gamma$. 

Finally, consider Eq.~\eqref{eq:constrC-evol} with $\kappa^\phi=0$. With the constraints ${\cal A}_i$ and ${\cal B}_i{}^j$ 
vanishing, this equation reduces to 
$\partial_t {\cal C}_i = {\mathcal L}_\beta {\cal C}_i$,
where ${\mathcal L}_\beta$ is the Lie derivative along the shift vector. It follows that the time evolution 
for ${\cal C}_i$ is simply a spatial diffeomorphism 
defined by the shift $\beta^i$. If initially the tensor components $C_i$ are given small nonzero values due to numerical error, 
these errors should stay small as long as the spatial coordinates remain well behaved. 

\subsection{Hyperbolicity} 
\label{sec:hyperbolicity}

The evolution equations~(\ref{gaugesector})--(\ref{nongaugesector}) form a quasilinear first order system,
\begin{equation}
\partial_t u = A(u)^i\partial_i u + F(u),
\label{Eq:QuasiLinearSystem}
\end{equation}
where the matrices $A^1$, $A^2$, $A^3$ and $F$ depend smoothly on the state vector $u =
(\alpha,\alpha_i,\beta^j,B^j,\beta_i{}^j,K,\phi,\phi_i,\tilde{\gamma}_{ij},\tilde{A}_{ij},\tilde{\gamma}_{kij},\tilde{\Lambda}^i)$. Such systems possess a local in time well-posed Cauchy problem if they are strongly hyperbolic, meaning that for each constant state vector $\uz$ in an appropriate open neighborhood and each normalized covector $n_i$ there exists a symmetric, positive definite matrix $H(\uz,n)$, depending smoothly on $\uz$ and $n_i$, such that $H(\uz,n) A(\uz)^i n_i$ is symmetric. The motivation for this definition stems from the principle of frozen coefficients~\cite{Kreiss89} in which the system~(\ref{Eq:QuasiLinearSystem}) is first linearized about some smooth solution $u$, and then its coefficients are frozen at a specific point $p$ of the spacetime manifold. Denoting by $\uz = u(p)$ the constant field that is obtained by freezing $u$ at $p$, and by $v$ the linearization of $u$, the system~(\ref{Eq:QuasiLinearSystem}) is then replaced by the linear, constant coefficient problem
\begin{equation}
\partial_t v = A(\uz)^i\partial_i v + {\cal F},
\label{Eq:FrozenCoeffSystem}
\end{equation}
with ${\cal F}$ some constant vector. When ${\cal F}=0$, this system describes the evolution of small amplitude, high-frequency perturbations of the quasilinear system~(\ref{Eq:QuasiLinearSystem}). Therefore, it is clear that a necessary condition for the well-posedness of the Cauchy problem for Eq.~(\ref{Eq:QuasiLinearSystem}) is that the principal parts of all frozen coefficient problems, {\em i.e.} Eq.~(\ref{Eq:FrozenCoeffSystem}) with ${\cal F} = 0$, lead to well-posed Cauchy problems. This turns out to be the case if and only if there exists a symmetric, positive definite matrix $H(\uz,n)$ such that $H(\uz,n) A(\uz)^i n_i$ is symmetric~\cite{Kreiss89}. Provided $H(\uz,n)$ depends smoothly on $\uz$ and $n$, the principle of frozen coefficients asserts that this is also a sufficient condition for the local in time well-posedness of the quasilinear problem~\cite{Kreiss89,Taylor99b}.

The existence of the ``symmetrizer'' matrix $H(\uz,n)$ implies, in particular, that the principal symbol $A(\uz,n) := A(\uz)^i n_i$ is diagonalizable and has a real spectrum for each $\uz$ and $n$. Once this necessary condition has been verified, the symmetrizer $H(\uz,n)$ can be constructed by diagonalizing $A(\uz,n) = S(\uz,n)\Lambda(\uz,n)S(\uz,n)^{-1}$ with $\Lambda(\uz,n)$ a real, diagonal matrix, and then setting $H(\uz,n) := (S(\uz,n)^{-1})^T S(\uz,n)^{-1}$. If $S(\uz,n)^{-1}$ depends smoothly on $\uz$ and $n$, this yields the required symmetrizer. The rows of $S(\uz,n)^{-1} u$ are the characteristic fields of the system~(\ref{Eq:QuasiLinearSystem}), and the diagonal entries of $\Lambda(\uz,n)$ are the corresponding characteristic speeds.

In our system~(\ref{gaugesector})--(\ref{nongaugesector}) the principle part naturally splits into two blocks, one of them, the ``gauge block'',  comes from the evolution equations~(\ref{gaugesector}) for the $20$ independent variables $\alpha$, $\alpha_i$, $\beta^j$, $B^j$, $\beta_i{}^j$, and $K$, and the other block, the ``non-gauge block'', comes from the evolution equations~(\ref{nongaugesector}) for the remaining variables. We first analyze the gauge block which is decoupled from the remaining block. Let us choose $\sigma = 1$. Through the replacements $\partial_t \mapsto \mu$ and $\partial_i \mapsto n_i$, we find that the eigenvalue problem $\mu v = A(\uz,n) v$ for this block reads
\begin{subequations}
\begin{eqnarray}
(\mu - \mathring\beta_n)\alpha &=& 0 \ ,
\\
(\mu - \mathring\beta_n)\alpha_i &=& -\mathring\alpha^2 \mathring f n_i K + \kappa^\alpha n_i\alpha \ ,\\
(\mu - \mathring\beta_n)\beta_j &=& 0 \ ,\\
(\mu - \mathring\beta_n)B_j &=& \mathring H\Bigl[ \beta_{nj} - \beta_{jn} \nonumber\\ & & 
        \quad{\ } + \frac{4}{3}\, n_j\beta_k{}^k - \frac{4\mathring\alpha}{3} n_j K \Bigr] \ ,\quad \\
(\mu - \mathring\beta_n)\beta_{ij} &=& \mathring\alpha^2 \mathring G n_i B_j + \kappa^\beta n_i\beta_j \ ,
\\
(\mu - \mathring\beta_n)K &=& -\alpha_n \ .
\end{eqnarray}
\end{subequations}
Here and in the following, the quantities $\mathring\alpha$, $\mathring\beta^i$, $\mathring\phi$, $\mathring\gamma_{ij}$ refer to the frozen lapse, shift, conformal factor and physical metric, respectively. Also, $\mathring f = f(\mathring\alpha,\mathring\phi)$ with similar definitions for $\mathring G$ and $\mathring H$. We assume that $\mathring f$, $\mathring G$, and $\mathring H$ are all positive. The covector $n_i$ is normalized such that $\mathring\gamma^{ij} n_i n_j = 1$. An index $n$ refers to contraction with $n^i = \mathring\gamma^{ij} n_j$; for example, $\alpha_n = n^i \alpha_i$. We have also used the frozen physical metric to lower indices: $\beta_i = \mathring\gamma_{ij}\beta^j$ and $\beta_{ij} = \beta_i{}^k\mathring\gamma_{kj}$.  

The characteristic fields and speeds for the gauge block are
\begin{widetext}
\begin{subequations}\label{charfieldsgaugeblock}
\begin{eqnarray}
& \beta_{AB} \ ,\ \alpha_A\ ,\ \beta_A \ ,\ \beta_{An} \ ,\ \alpha \ ,\ \beta_n \ ,
& \mu = \mathring\beta_n \ ,\\
& G^{(\pm)}_A \equiv B_A + \frac{\kappa^\beta}{\mathring\alpha^2 \mathring G} \beta_A
\pm \frac{1}{\mathring\alpha}\sqrt{\frac{\mathring H}{\mathring G}}(\beta_{nA} - \beta_{An}) \ ,
& \mu = \mathring\beta_n \pm\mathring\alpha\sqrt{\mathring G \mathring H} \ ,\\
& G^{(\alpha,\pm)} \equiv K - \frac{\kappa^\alpha}{\mathring\alpha^2 \mathring f}\alpha
  \mp \frac{1}{\mathring\alpha\sqrt{\mathring f}} \alpha_n\ ,
& \mu = \mathring\beta_n \pm \mathring\alpha\sqrt{\mathring f} \ ,\\
& G^{(\beta,\pm)} \equiv B_n + \frac{\kappa^\beta}{\mathring\alpha^2 \mathring G} \beta_n
 \pm \frac{\lambda}{\mathring\alpha \mathring G}\mathring\gamma^{rs}\beta_{rs}
 \mp \frac{4\mathring H}{3(\lambda^2 - \mathring f)}\left( \lambda K 
 - \frac{\kappa^\alpha}{\mathring\alpha^2\lambda}\alpha 
 \mp \frac{1}{\mathring\alpha} \alpha_n
\right) \ ,\quad
& \mu = \mathring\beta_n \pm \mathring\alpha\lambda \ .
\end{eqnarray}
\end{subequations}
\end{widetext}
Here, indices $A$ and $B$ refer to contraction with unit vectors orthogonal to
$n^i$, and we have set $\lambda \equiv \sqrt{4\mathring G\mathring H/3}$. The
characteristic fields are well--defined and independent from each other
as long as $\lambda^2 \neq \mathring f$. This restriction on hyperbolicity, which is more explicity written as 
\begin{equation}
        4\mathring G \mathring H \neq 3\mathring f \label{eq:sh_gauge_condition}\, , 
\end{equation} 
is also required for strong hyperbolicity in the second order BSSN system \cite{Beyer:2004sv}.  

The eigenvalue problem $\mu v = A(\uz,n) v$ for the non-gauge block is given by
\begin{widetext}
\begin{subequations}\label{Eq:NGBlock}
\begin{eqnarray}
(\mu - \mathring\beta_n)\phi &=& 0 \ ,\\
(\mu - \mathring\beta_n)\phi_i &=& -\frac{\mathring\alpha}{6} n_i K + \frac{1}{6} n_i\beta_k{}^k + \kappa^\phi n_i\phi \ ,\\
(\mu - \mathring\beta_n)\tilde{\gamma}_{ij} &=& 0 \ ,\\
(\mu - \mathring\beta_n)\tilde{A}_{ij} &=& -\frac{\mathring\alpha}{2}\left[ \tilde{\gamma}_{nij} \right]^{TF}
 + e^{-4\mathring\phi}\left[ \mathring\alpha n_{(i}\tilde{\Lambda}_{j)}
 - 2\mathring\alpha n_{(i} \phi_{j)} - n_{(i} \alpha_{j)} \right]^{TF} \ ,\\
(\mu - \mathring\beta_n)\tilde{\gamma}_{kij} &=& -2\mathring\alpha n_k\tilde{A}_{ij}
 + 2e^{-4\mathring\phi} n_k \left[ \beta_{(ij)} \right]^{TF}
 + \kappa^\gamma n_k\tilde{\gamma}_{ij}\ ,
\label{eq:mugammakij}\\
(\mu - \mathring\beta_n)\tilde{\Lambda}_j &=& \beta_{nj} - \beta_{jn} + \frac{4}{3} n_j\beta_k{}^k
 - \frac{4\mathring\alpha}{3}\, n_j K \ ,
\end{eqnarray}
\end{subequations}
\end{widetext}
where $\tilde{\Lambda}_k = e^{-4\mathring\phi}\mathring\gamma_{k\ell}\tilde{\Lambda}^\ell$. 
This block consists of 32 independent variables: 1 variable $\phi$, 3 variables $\phi_i$, 5 variables $\tilde\gamma_{ij}$ (since $\tilde{\gamma}_{ij}$ is the linearization of a metric with fixed determinant), 5 variables $\tilde A_{ij}$ (since $\tilde A_{ij}$ is symmetric and trace--free), 15 variables $\tilde \gamma_{kij}$ (since $\tilde \gamma_{kij}$ is symmetric and trace--free in $i$ and $j$), and 3 variables $\tilde\Lambda_j$. 

The non--gauge block needs $\kappa^\phi = 0$ to be
diagonalizable. With $\kappa^\phi = 0$ the characteristic fields and speeds are 
\begin{widetext}
\begin{subequations}
\begin{eqnarray}
& \phi \ ,\ 
  Z_0 \equiv \phi_n - \frac{1}{8}\tilde{\Lambda}_n \ ,\ 
  \phi_A \ ,\ 
  \tilde{\gamma}_{ij} \ ,\ 
  Z_i\equiv \mathring H\tilde{\Lambda}_i - B_i \ ,\ 
  \tilde\gamma_{Aij} \ , 
& \mu = \mathring\beta_n\ ,\\
& {V}_{AB}^{(\pm)} \equiv \tilde{A}_{AB}^{tf}
 - \frac{\kappa^\gamma}{2\mathring\alpha}\tilde{\gamma}_{AB}^{tf}
 - \frac{1}{\mathring\alpha} e^{-4\mathring\phi}{\beta}_{(AB)}^{tf}
 \mp \frac{1}{2}\, \tilde{\gamma}_{nAB}^{tf} \ ,
& \mu = \mathring\beta_n \pm\mathring\alpha \ ,\\
& V_{nA}^{(\pm)} \equiv \tilde{A}_{nA}
 - \frac{\kappa^\gamma}{2\mathring\alpha} \tilde{\gamma}_{nA}
 - \frac{1}{\mathring\alpha} e^{-4\mathring\phi}\beta_{An}
 \mp \frac{1}{2}\left[ \tilde{\gamma}_{nnA}
  - e^{-4\mathring\phi}\left( \tilde{\Lambda}_A - 2 \phi_A 
  - \frac{1}{\mathring\alpha}\alpha_A \right) \right] \ ,
& \mu = \mathring\beta_n \pm \mathring\alpha \ ,\\
& V_{nn}^{(\pm)} \equiv \tilde{A}_{nn} 
 - \frac{\kappa^\gamma}{2\mathring\alpha}\tilde{\gamma}_{nn}
 + \frac{1}{\mathring\alpha} e^{-4\mathring\phi}\beta_{AB}\delta^{AB}
 - \frac{2}{3} e^{-4\mathring\phi} K 
 \mp \left[ \frac{1}{2}\tilde{\gamma}_{nnn}
 - \frac{2}{3} e^{-4\mathring\phi}\left( \tilde{\Lambda}_n - 2 \phi_n \right)
 \right] \ ,\quad
& \mu = \mathring\beta_n \pm\mathring\alpha \ ,\qquad
\label{Eq:Vnn}
\end{eqnarray}
\end{subequations}
\end{widetext}
where the superscript $tf$ refers to the trace--free part in the transverse directions; for instance, $\tilde{A}_{AB}^{tf} = \tilde{A}_{AB} - \frac{1}{2}\delta_{AB}\delta^{CD}\tilde{A}_{CD}$. 

Provided the functions $f$, $G$ and $H$ depend smoothly on $(\alpha,\phi)$ and satisfy the restriction~(\ref{eq:sh_gauge_condition}), a smooth symmetrizer $H(\uz,n)$ can be constructed from the characteristic fields as described at the beginning of this subsection.

In the analysis above, we have assumed that all the algebraic constraints are identically satisfied, which is consistent with the evolution equations after the modifications to Eq.~(\ref{eq:dkijtilde-evol}) described in Section~\ref{SubSec:CP}.\footnote{These modifications do not change the principal part of the equations when the algebraic constraints hold.} If none of the algebraic constraints are enforced, then Eq.~(\ref{Eq:NGBlock}) yields $(\mu - \mathring\beta_n)(\mathring{\gamma}^{ij}\tilde{\gamma}_{ij}) = 0$, $(\mu - \mathring\beta_n)(\mathring{\gamma}^{ij}\tilde{A}_{ij}) = 0$ and $(\mu - \mathring\beta_n)(\mathring{\gamma}^{ij}\tilde{\gamma}_{kij}) = n_k[-2\mathring{\alpha}(\mathring{\gamma}^{ij}\tilde{A}_{ij}) + \kappa^\gamma(\mathring{\gamma}^{ij}\tilde{\gamma}_{ij})]$ which is a weakly hyperbolic system. In order to obtain a strongly hyperbolic system one could enforce only the trace constraint $\tilde{A}_i{}^i = 0$ and replace ${\cal D}_{kij}$ by its trace-free part over $ij$ in the right-hand side of Eq.~(\ref{eq:dkijtilde-evol}). In this case, $\mathring{\gamma}^{ij}\tilde{\gamma}_{ij}$ and $\mathring{\gamma}^{ij}\tilde{\gamma}_{kij}$ are characteristic fields with speeds $\mu = \mathring\beta_n$ and one has to perform the replacements
\begin{displaymath}
\tilde{\gamma}_{nn} \mapsto \frac{2}{3}\left( \tilde{\gamma}_{nn} 
 - \frac{1}{2}\tilde{\gamma}_{AB}\delta^{AB} \right)
\end{displaymath}
and
\begin{displaymath}
\tilde{\gamma}_{nnn} \mapsto \frac{2}{3}\left( \tilde{\gamma}_{nnn} 
 - \frac{1}{2}\tilde{\gamma}_{nAB}\delta^{AB} \right)
\end{displaymath}
in the expression for $V_{nn}^{(\pm)}$ in Eq.~(\ref{Eq:Vnn}).

\section{Numerical Experiments} \label{sec:Results}

Here we summarize results of numerical experiments of the first order 
BSSN formulation described in the previous sections, with different
numerical approaches and codes, from more traditional ones for which
there is more experience (finite differences with adaptive mesh
refinement), to a promising approach that only recently is making
its way into numerical relativity (Discontinuous Galerkin Finite
Elements, restricted
here to spherical symmetry). 

In more detail, our approach and summary of numerical experiments with the FOBSSN formulation is in the following order: 
\begin{enumerate}
\item Section~\ref{sec:ResultsFD}: Two Apples-with-Apples tests
  \cite{Alcubierre2003:mexico-I}, as well as result from single and
  binary black hole moving puncture simulations using finite
  differences with AMR\@. Most of our simulations show no signs of time or numerical
  instabilities. By time-stability in time dependent problems it is referred to the numerical solution not growing at any fixed resolution in time unless the exact solution does so. Numerical stability refers to the property that at any fixed time the errors in the numerical solution decrease with increasing resolution. In our simulations we have found the solution to be both time and numerically stable. 
 
The non-linear gauge wave test with a large amplitudes shows a global time-instability (yet not a numerical one) that is expected; see \cite{Alcubierre2003:mexico-I}. This suggests that the addition of the extra constraints present when enlarging the system to a purely first order formulation does not trigger any obvious instability. The extracted gravitational waves are found to be consistent with simulations done using the standard second-order BSSN formulation. In addition, the FOBSSN results are often more accurate than BSSN results using the same resolution.

\item Section~\ref{sec:ResultsDG}:  With the standard BSSN gauge conditions a non-rotating black hole is driven to the trumpet solution \cite{Hannam:2009ib}. However, this would require either the moving punctures technique \cite{Campanelli:2005dd,Baker:2005vv} or the turduckening one \cite{Brown:2007pg,Brown:2008sb,Etienne:2007hr}. In the first case the equations become singular at the puncture locations, which would be difficult to deal with using a very high order method such as those motivating the current paper. The turducken approach, on the other hand,  smooths the solution inside the black hole while guaranteeing that the associated constraint violations do not ``leak'' to the exterior of the black hole.  As a first step in that direction we test a discontinuous Galerkin (dG) approach using the FOBSSN system for black holes in spherical symmetry, first using excision. 
\item Section~\ref{sec:turducken}: As a final step, we perform turducken black hole dG simulations in spherical symmetry, both using FOBSSN and the standard second order formulation. We are able to perform long term and stable evolutions with the standard second BSSN formulation, but find numerical instabilities with FOBSSN.
\end{enumerate}

Discussions about all these experiments, their interpretation, and proposed next steps are discussed in Section~\ref{sec:Conclusions}.  Next we provide a somewhat detailed summary of these numerical experiments.

\subsection{Finite differences} \label{sec:ResultsFD}
We have implemented the first-order system
(\ref{gaugesector})--(\ref{nongaugesector})
using the \texttt{Cactus}
framework~\cite{Goodale:2002a, Cactuscode:web}, and employing the
\texttt{Carpet} adaptive mesh refinement (AMR)
driver~\cite{Schnetter-etal-03b, CarpetCode:web}. We used the
Mathematica package \texttt{Kranc}~\cite{Husa:2004ip, Kranc:web} to
expand the FOBSSN equations to C code, in the same manner as already
for the \texttt{McLachlan} code~\cite{Brown:2008sb, McLachlan:web}.
Both the Mathematica notebook as well as the resulting C code will be
made available for public download as part of the Einstein Toolkit
\cite{Loffler:2011ay, EinsteinToolkit:web} under the name
\texttt{Carlile}.

Our implementation supports arbitrary finite differencing orders and
time integration orders; below, we use fourth order accurate stencils
and a fourth order Runge-Kutta time integrator. We use fifth order
Kreiss-Oliger dissipation as well as fifth order spatial interpolation
at AMR boundaries. We use buffer zones and tapered grids
\cite{Brown:2008sb} to avoid time interpolation at mesh refinement
boundaries. This makes all simulations fully fourth order convergent.
The algebraic constraints $\tilde\gamma^{ij}\tilde A_{ij}=0$ and
$\tilde\gamma^{ij} \tilde\gamma_{kij}=0$ are enforced every time the state
vector is modified. However, $\tilde\gamma=1 = \overline\gamma$ is not
enforced, but is nevertheless assumed to hold throughout the
implementation. Our constraint damping and related parameter settings
are listed in table \ref{tab:constr-param}.
We impose simple outgoing radiation (Sommerfeld)
boundary conditions on all fields.

\begin{table}
  \begin{tabular}{lcl|ll}
    Code Name & Symbol & Eq. & Value & Comment \\\hline
    
    harmonicN & $\alpha^2 f$ & (\ref{eq:alpha-evol}) & 1 & ($1+\log$)
    \\
    
    harmonicF & $\alpha^2 f$ & (\ref{eq:alpha-evol}) & 2.0 &
    ($1+\log$) \\
    
    ShiftGammaCoeff & $\alpha^2 G$ & (\ref{eq:beta-evol}) & 0.75 &
    (std.\ choice) \\
    
    BetaDriver & $\eta$ & (\ref{eq:B-evol}) & 1.0 \\
    
    DAlphaDriver & $\kappa^\alpha$ & (\ref{eq:alpha-evol}) & 1.0 \\
    
    DBetaDriver & $\kappa^\beta$ & (\ref{eq:betai-evol}) & 1.0 \\
    
    DphiDriver & $\kappa^\phi$ & (\ref{eq:phi-evol}) & 0.0 & (not
    enforced) \\
    
    DgtDriver & $\kappa^\gamma$ & (\ref{eq:dkijtilde-evol}) & 1.0 \\
    
    sigma & $\sigma$ & (\ref{eq:gamma-evol}) & 1.0 \\
  \end{tabular}
  \caption{Constraint damping and related parameter settings}
  \label{tab:constr-param}
\end{table}

\paragraph{Robust Stability Test.}

\begin{figure}
  \includegraphics[width=0.47\textwidth]{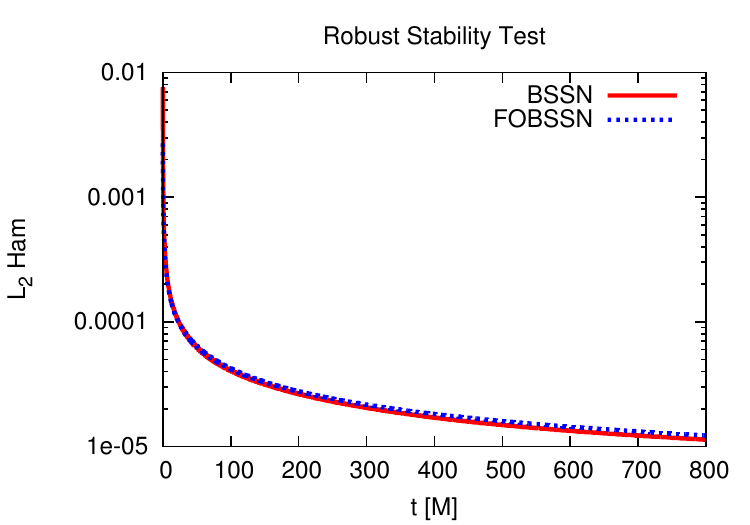}
  \caption{Robust stability test comparing of BSSN and FOBSSN\@. A
    cubic domain is initialised with Minkowski data and a low level of
    noise in all variables, and then evolved with periodic boundary
    conditions. This tests linear stability of the formulation. The
    fact that the constraint violation decreases indicates stability.
    Both BSSN and FOBSSN perform very similarly here.}
  \label{fig:sbh_MAH}
\end{figure}

One of the most important and most fundamental test for a formulation
of the Einstein equations and its numerical implementation is a
\emph{robust stability test}, which can demonstrate linear stability.
The simulation domain is initialised with Minkowski data plus a small
amount of noise, and then let to evolve freely
\cite{Alcubierre2003:mexico-I}. Here, we use a cubic domain with
$40^3$ grid points and periodic boundary conditions, and a noise
amplitude of $A=10^{-6}$ in all BSSN or FOBSSN variables. Figure
\ref{fig:sbh_MAH} compares the performance of BSSN and FOBSSN, and
finds very similar behaviour. In particular, the $L_2$ norm of the
Hamiltonian constraint decreases steadily over time, indicating robust
stability.

\paragraph{Nonlinear Gauge Wave.}

A very demanding test is evolving a nonlinear gauge wave. This is a
fully nonlinear solution of the Einstein equations where the exact
solution is known, as it is a flat spacetime in a complex,
time-dependent coordinate system \cite{Alcubierre2003:mexico-I}. Here,
we use a one-dimensional domain with $40\times 1\times 1$ grid points with
periodic boundaries, and evolve with the full 3D formulation. We test
two cases, a large amplitude ($A=0.1$) and a small amplitude
($A=0.01$), employing the $\exp\sin$ form of the gauge wave.

\begin{figure}
  \includegraphics[width=0.47\textwidth]{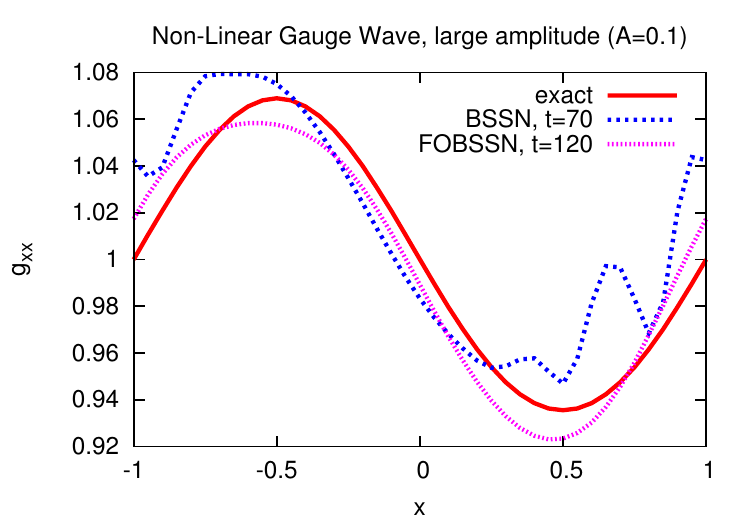}
  \caption{Nonlinear gauge wave test, $A=0.1$ (large
    amplitude), comparing of BSSN and FOBSSN\@. At $t=70$ (35 crossing
    times), the BSSN solution has broken down (become irregular) due
    to accumulation of numerical errors. The FOBSSN breaks down much
    later, shortly after $t=120$ (60 crossing times); at $t=120$, the
    FOBSSN solutions is still regular, and has only picked up a phase
    error and a global downwards drift in the metric.}
  \label{fig:gw_gt11}
\end{figure}

Figure \ref{fig:gw_gt11} shows results from the large-amplitude case.
This is a very demanding case that is known to go unstable quickly for
many formuations of the Einstein equations
\cite{Alcubierre2003:mexico-I}. Here we observe that both the
evolutions with BSSN and the FOBSSN formulations break down; however,
the FOBSSN evolution lasts for about twice as many crossing times. We
also observe that the breakdown mechanisms for BSSN and FOBSSN are
different -- the BSSN result develops high-frequency noise
(depicted), while in the FOBSSN result the metric drifts downwards,
i.e.\ the proper size of the simulation domain decreases.

\begin{figure}
  \includegraphics[width=0.47\textwidth]{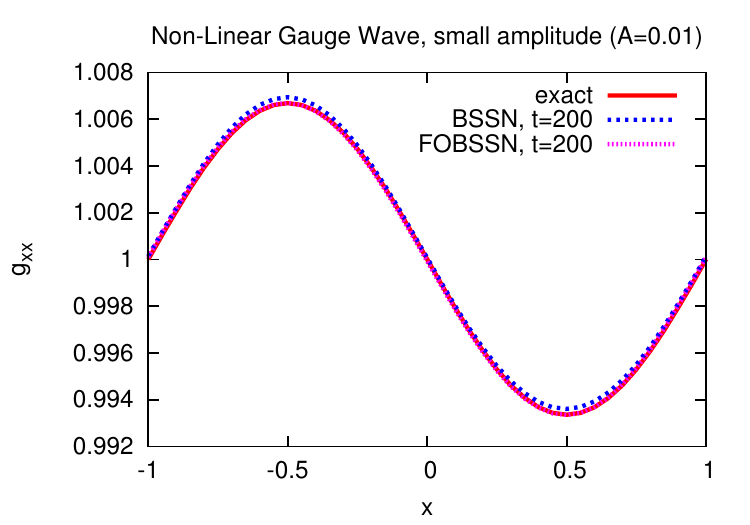}
  \caption{Nonlinear gauge wave test, $A=0.01$ (small
    amplitude), comparing of BSSN and FOBSSN\@. At $t=200$ (100
    crossing times), both the BSSN and FOBSSN solutions are still
    fine. However, the BSSN solution has begin to drift upwards much
    more than the FOBSSN solution.}
  \label{fig:gw-small_gt11}
\end{figure}

Figure \ref{fig:gw-small_gt11} shows results from the small-amplitude
case. This is a less demanding case where most formulations of the
Einstein equations can perform long-term evolutions
\cite{Alcubierre2003:mexico-I}. After 100 crossing times, both the
BSSN and FOBSSN result looks fine; however, the BSSN result exhibits a
much larger upwards drift in the metric.

\paragraph{Single Puncture Black Hole.}

A much more interesting test of the FOBSSN formulation is evolving a
puncture black hole. Here we choose a rotating puncture with total
mass $M=1$ and spin $a=0.7$, set up via the \texttt{TwoPunctures}
thorn \cite{Ansorg:2004ds}. These initial conditions are conformally
flat and contain some gravitational radiation, and the black hole is
expected to relax to a stationary state after some time. In the
figures below, we use a length unit $M$ that corresponds approximately
to the ADM mass of the system, which is $M_{ADM} = 1.00252\,M$. The
black hole horizon has a coordinate radius of approximately $0.376$ 
initially and $0.766$ at late times.

We employ eight levels of mesh refinement in a cubic domain, placing
refinement boundaries at $x=[1,2,4,8,16,64,128]\,M$, and placing the
outer boundary at $258.048\,M$. The resolution on the finest level,
which encompasses the horizon at all times, is $h=0.032\,M$.

\begin{figure}
  \includegraphics[width=0.47\textwidth]{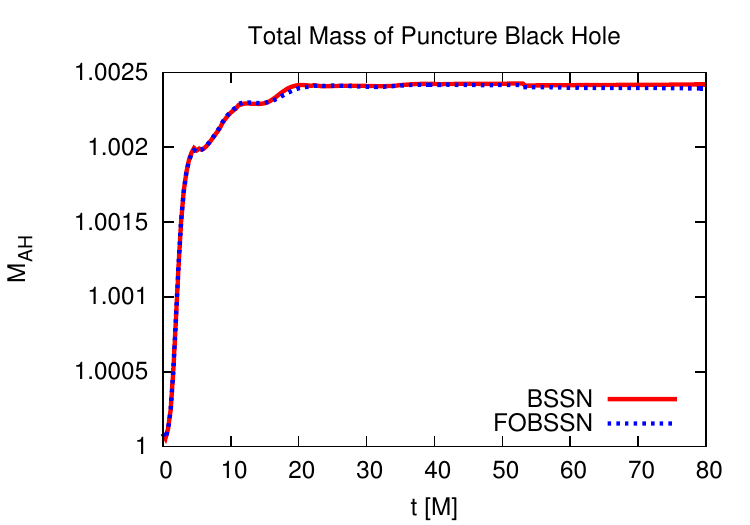}
  \caption{Black hole total mass vs.\ time for a single puncture black
    hole with spin $a=0.7$, comparing the accuracy of BSSN and
    FOBSSN\@. This is an initially non-stationary solution that
    evolves towards a trumpet solution. BSSN and FOBSSN perform very
    similarly here, and in particular, the moving
    puncture/turduckening approach to singularity handling seems to
    work fine for FOBSSN\@.}
  \label{fig:sbh_M}
\end{figure}

\begin{figure}
  \includegraphics[width=0.47\textwidth]{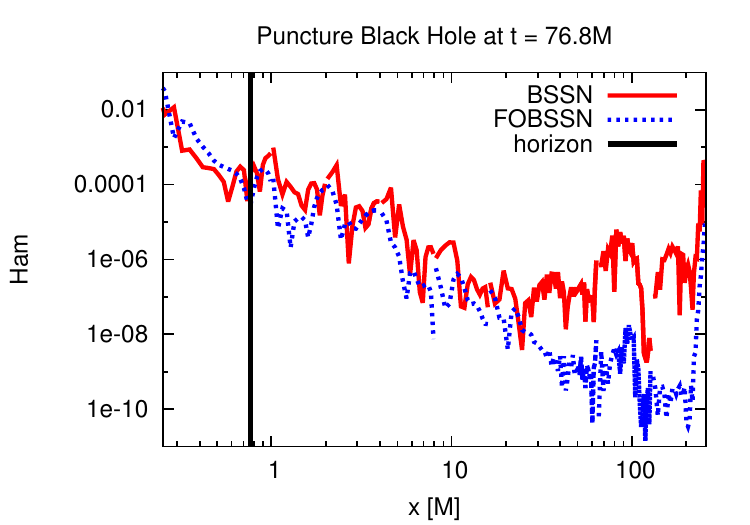}
  \caption{Hamiltonian constraint violation along the $x$ axis at
    $t=76.8\,M$ for a single puncture black hole, comparing the
    accuracy of BSSN and FOBSSN\@. The constraint violation increases
    towards the black hole (located at $x=0$), where the horizon has a
    coordinate radius of about $r=0.766\,M$ at this time. The
    constraint violation near $x=100\,M$ is caused by outer boundary
    effects. FOBSSN seems to perform slightly better than BSSN
    in the bulk of the domain,
    and significantly better near the outer boundary.}
  \label{fig:sbh_Ham}
\end{figure}

Figure \ref{fig:sbh_M} shows the total mass of the black hole as
calculated by the \texttt{QuasiLocalMeasures} thorn
\cite{Dreyer:2002mx}. After an initial transient lasting about
$20\,M$, the spacetime becomes manifestly stationary. The angular
momentum (not shown) remains approximately constant at
$J=0.701\pm0.006\,M^2$. Figure \ref{fig:sbh_Ham} shows a snapshot of
the Hamiltonian constraint in this stationary state along the $x$ axis
at $t=76.8M$. As expected, the constraint violation increases towards
the black hole. Both BSSN and FOBSSN perform approximately the same
except near the outer boundary, where FOBSSN seems superior.

\paragraph{Inspiralling Binary Black Holes.}

As a more advanced test, we also evolve inspiralling binary black
holes, using the \emph{R1} configuration of \cite{Baker:2006yw,
  Brugmann:2008zz}. This configuration performs about $1.8$ orbits
prior to merger, with a common apparent horizon first found at roughly
$t=160\,M$, where the ADM mass $M_{ADM}=0.966\,M$ sets the scale. The
initial individual black holes have masses $M_1=M_2=0.505\,M$ and have
no spin.

We use $9$ levels of adaptive mesh refinement and placed the outer
boundary at $320\,M$. The simulations were performed at two
resolutions of $h=M/28.8$ and $M=1/38.4$, where $h$ denotes the
gridspacing on the finest grid, see also \cite{Baker:2006yw,
  Brugmann:2008zz} for comparison. We use fourth order accurate finite
differencing stencils with lop-sided stencils for advection
terms~\cite{Brown07}
and fourth order Runge Kutta time integration with Berger-Oliger
sub-cycling in time. We do not employ tapered grids, using second
order time interpolation where necessary on mesh refinement
boundaries.

Due to the larger number of constraints in the first order
formulation, one would expect a better accuracy in the second order
formulation for the same number of grid points \cite{HinderThesis2005,
  Calabrese:2005ft, Chirvasa:2008xx, Kreiss:2010aa}. That seems indeed
to be the case here: we find that the BSSN formulation allows us to
use a lower resolution than the FOBSSN formulation to achieve
time-stability.

\begin{figure}
  \includegraphics[width=0.47\textwidth]{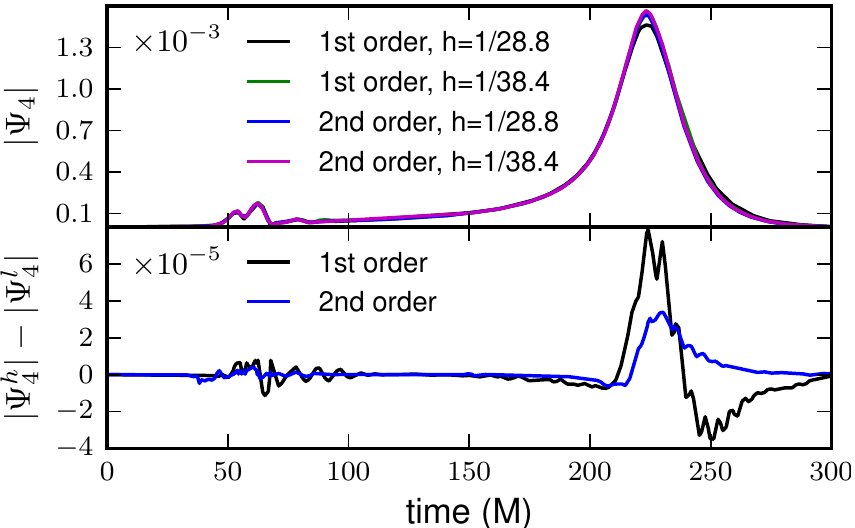}
  \caption{Comparison between simulations of the standard first order
    in time, second order in space implementation of BSSN and a fully
    first order reduction for a binary black hole inspiral system. The
    top panel shows the amplitude of the $\ell=2$, $m=2$ mode of
    $\Psi_4$ extracted at $r=50$ at two resolutions $h$ for both
    implementations. The bottom panel shows the difference between
    different resolutions.}
  \label{fig:bssn_inspiral}
\end{figure}

Figure \ref{fig:bssn_inspiral} shows the amplitude of the $\ell=2$,
$m=2$ mode of the Weyl scalar $\Psi_4$, extracted on a coordinate
sphere with radius $r=50\,M$. The low-resolution FOBSSN simulation is
visibly different from the other simulations at the peak of the
amplitude. However, the high resolution FOBSSN simulation agrees with
both BSSN resolutions.

Throughout the simulations, there is generally a good agreement
between all runs. The lower panel of figure \ref{fig:bssn_inspiral}
shows the difference in the amplitude between high- and low-resolution
runs for the BSSN and FOBSSN formulations, indicating that BSSN may have
a smaller relative error.

\subsection{Discontinuous Galerkin} \label{sec:ResultsDG}

Next we consider a dG scheme for the spherically reduced first order BSSN system (see \cite{Field:2010mn} for a dG implementation of the second order form of the BSSN equations). There are some important differences with FOBSSN in three dimensions, which arise when specializing to spherical symmetry. First, the constraint $\tilde{A}^i_i = 0$ is exactly satisfied by virtue of the spherically symmetric restriction. Second, terms proportional to $\sigma$ in Eq.~(\ref{eq:gamma-evol}) are identically zero and so we set $\sigma = 0$. Finally, spherical symmetry is no longer associated with the obvious choice $\overline\Gamma^i_{jk} = 0$ and $\overline{\gamma}=1$. As a consequence, using the fiducial covariant derivative will give rise to terms which feature $\overline \gamma$ and its derivatives. Our approach is to notice the fact that the covariant divergence $\overline D_i \beta^i$ only depends on $\overline \gamma$ and use the constraint $\tilde\gamma  - \overline\gamma = 0$ to replace $\overline D_i \beta^i \rightarrow \tilde D_i \beta^i$ throughout BSSN system (\ref{eq:evolBSSN}). Furthermore, we use $\tilde \Gamma^r$ in place of $\tilde \Lambda^r$, which results in a strongly hyperbolic spherically reduced BSSN system, explicitly given by Eq.~(10) of~\cite{Field:2010mn}. For a complete discussion see Ref.~\cite{Brown:2007nt}.

We have discretized the first-order spherically reduced system with a nodal dG method \cite{Hesthaven_DG_book,Cockburn1998199}. Similar to a multi-domain pseudo-spectral collocation method,  a dG approach provides for a multi-domain treatment of the geometry where the numerical solution on each subdomain is given by a (time-dependent) polynomial of arbitrarily high order degree $N$. On every subdomain each component of the PDE is required  to be satisfied in a suitable weak (integral) sense, yielding $(N+1)$ ordinary differential equations often known as Galerkin conditions.  
Adjacent subdomains are coupled in a stable manner through a suitable numerical flux term~\cite{Hesthaven_DG_book}. The resulting scheme is nearly identical to the one presented in~\cite{Field:2010mn} with the notable exception of the absence of second order operators. Hence, we use the standard local Lax--Friedrichs form for the numerical flux~\cite{Hesthaven_DG_book}, while in the second-order system, to which we will sometimes compare, a penalized central flux is used for a stable treatment second order operators (see page 13 of Ref.~\cite{Field:2010mn} for more details). The integration in time is implemented using the method-of-lines with a fourth order Runge-Kutta scheme. After each timestep an exponential filter is applied to the top two-thirds of the modal coefficients to control alias driven instabilities. Furthermore, in our dG implementation of both the second and first order BSSN system we have empirically observed that the conformal metric coefficients must {\em not} be filtered otherwise the scheme becomes unstable. A perhaps related observation is that enforcing the constraint $\tilde \gamma = \overline \gamma$  triggers an instability at very early times. Neither this constraint nor the spherically symmetric version of Eq.~(\ref{tracegammakij}) are enforced in our dG implementation.

All simulations presented next are for the Schwarzschild metric in
conformal ingoing Kerr-Schild coordinates\footnote{In these
  coordinates, at least initially, $\widetilde \gamma_{ij} = {\rm diag}( 1, r^2, r^2 \sin^2\theta)$ and so the algebraic constraint is $\widetilde \gamma = \overline\gamma = r^4 \sin^2\theta$.}~\cite{Field:2010mn}. The source terms $S_\alpha$, $S_\beta^r$, and $S_B^r$ in the gauge equations (\ref{eq:bm-slicing}) and (\ref{gammadrivershifteqns}) are chosen so that the numerical solution is time-independent. Typical choices for $f$, $G$, $\eta$ and $\sigma$ are used; in detail: $f=2/\alpha$, $G=3/4\alpha^{-2}$, $\eta=50$, and terms proportional to $\sigma$ are identically zero. Furthermore, we set\footnote{The dG code evolves the conformal factor $\chi = e^{- 4 \phi}$. Nevertheless, we continue to refer to the conformal factor as $\phi$ in this section.} $H = e^{4\phi} / L$ and choose $L=10$ such that the excision surface is not too close to $r=0$, where field gradients are large. All damping parameters $\kappa^\alpha$, $\kappa^\beta$, $\kappa^\phi$ and $\kappa^\gamma$ are set to $20$. We find that different values of $L$ have negligible effect on the scheme's stability, and in particular no dependence on the location of the $e^{4\phi} = 2\alpha L$ surface of potential weak hyperbolicity [cf. Eq.~(\ref{eq:sh_gauge_condition})].

\begin{figure}
 \begin{center}
 \includegraphics[width=3.4in]{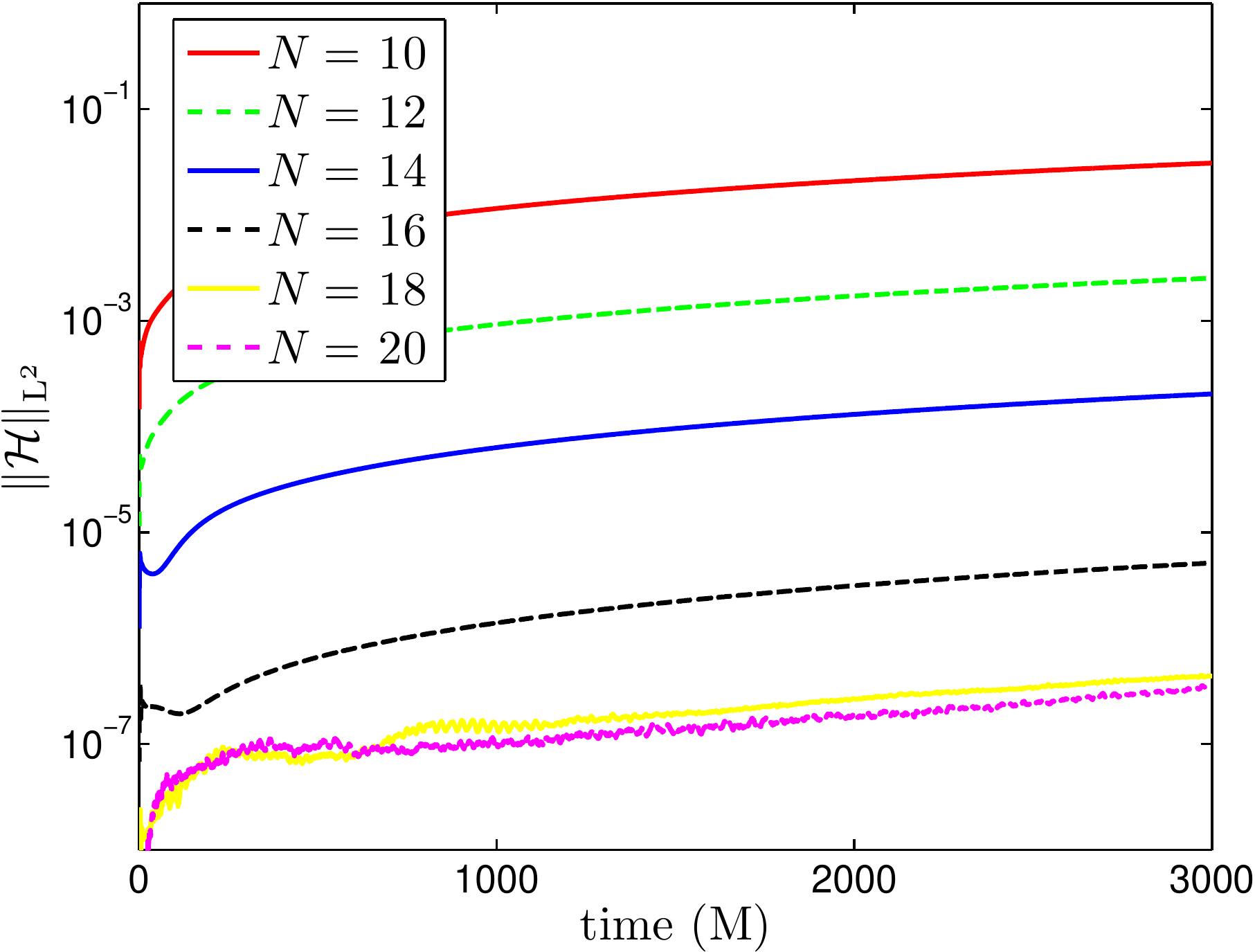}
 \end{center}
 \caption{Discontinuous Galerkin evolutions of a black hole in spherical symmetry, using excision. See the text in Section \ref{sec:ResultsDG} for more details. The last two resolutions have essentially reached double precision roundoff errors. }
 \label{fig:FOBSSN_boundary_fixedR50}
 \end{figure}

\begin{figure}
 \begin{center}
 \includegraphics[width=3.4in]{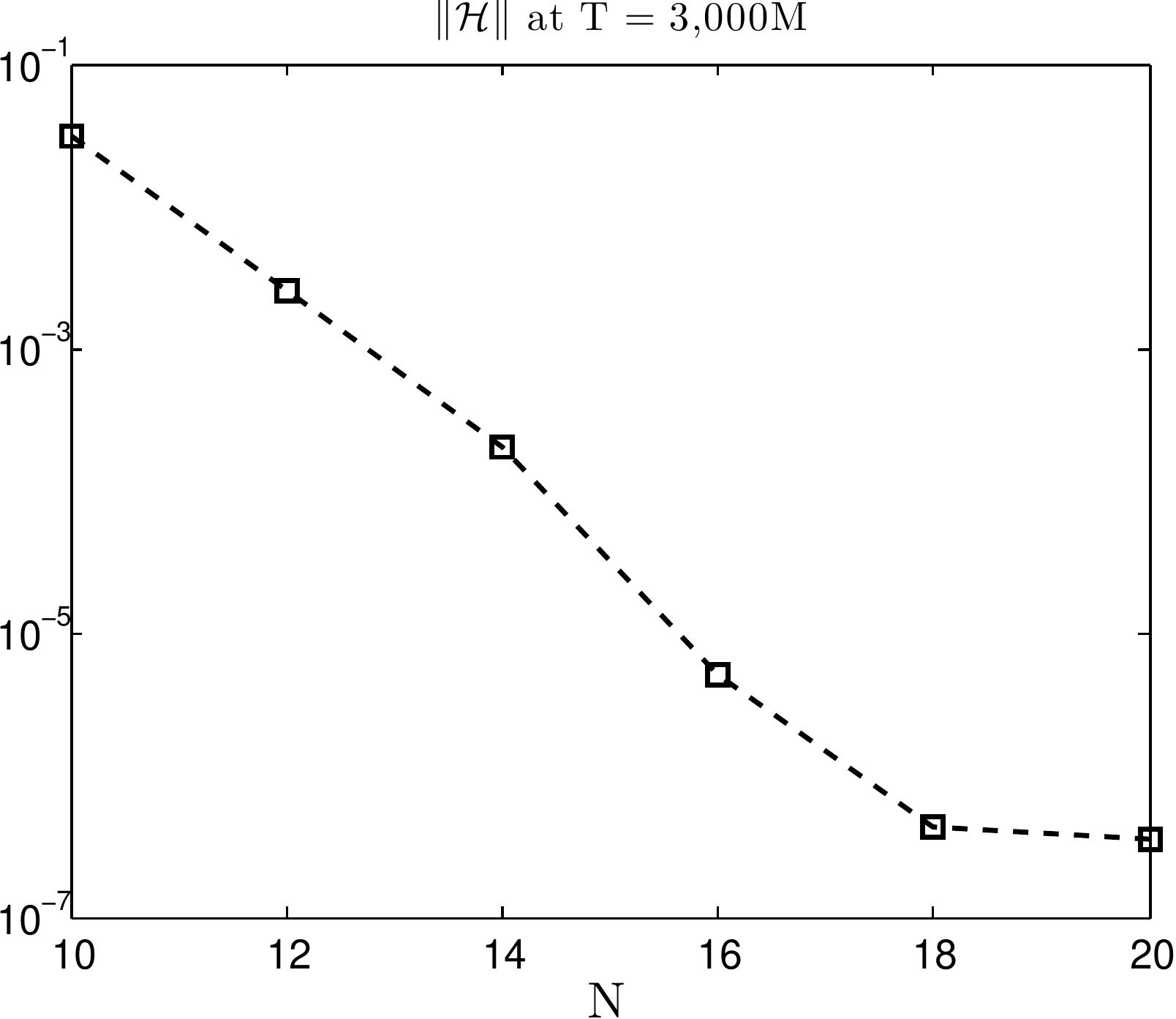}
 \end{center}
 \caption{Exponential convergence of Discontinuous Galerkin evolutions with polynomial order, see Section \ref{sec:ResultsDG}. The norm used is the $L_2$ one. }
 \label{fig:convergenceDGFOBSSN}
 \end{figure}

The radial domain $[0.4, 50]M$ is covered by 100 equally sized subdomains\footnote{This is far from optimal, since a better choice would be to have the size of the domains increase with radius. However, it suffices to make our point about stability and convergence.}.  We treat the inner boundary by excision. At the outer physical boundary we specify the analytic values for the incoming characteristic modes,  which, for the spherically reduced system considered here, are given by Eqs.~(17a-i) listed in~\cite{Field:2010mn} (Dirichlet conditions). Figs.~\ref{fig:FOBSSN_boundary_fixedR50} and~\ref{fig:convergenceDGFOBSSN} show that the scheme converges exponentially with $N$ and is able to achieve very long run times. In an attempt to remove the slow growth in time for any fixed resolution seen in the Hamiltonian constraint we varied our numerical setup including the exponential filter parameters (the number of filtered modes, the dissipation exponent, and which variables to filter), the timestep, the $B^r$ damping parameter $\eta$, the numerical flux dissipation parameter, and the auxiliary field damping parameters, without significant improvements. 

\subsection{Turduckening} \label{sec:turducken}

Successful numerical evolution of binary black hole systems require a suitable treatment of singularities. There are three distinct techniques used:  moving-punctures~\cite{Campanelli:2005dd,Baker:2005vv}, excision, and smoothing via turduckening~\cite{Etienne:2007hr,Brown:2007pg,Brown:2008sb}. State-of-the-art second order BSSN codes avoid the complications of excision, which require horizon tracking.  A moving-puncture technique was used in our finite difference implementation in Section \ref{sec:ResultsFD}, while the dG code in Section~\ref{sec:turducken}) each relied on excision.  It has been shown that the usual gauge conditions are attractors of the trumpet solution~\cite{Waxenegger:2011ci,Witek:2010es,Thierfelder:2010dv,Hannam:2008sg} for which there is an incoming characteristic mode {\em even at the puncture}~\cite{Brown:2009ki} -- generically we therefore do not expect an excision surface where no boundary conditions are required to exist. As the majority of BSSN implementations without excision have been thus far limited to finite difference methods, one wonders how other methods might deal with singularities.  In this subsection we give a preliminary look at turduckening for the nodal dG code. 

We follow the turduckening technique described in Ref.~\cite{Brown:2008sb}. Singular initial data in the interior of the black hole is replaced with smooth constraint violating data. The prescription for such smoothing used here is as follows. If the computational domain is $r \in [0,R_\mathrm{max}/M]$, we select a coordinate location $r_t$ inside the horizon and make the replacement $r \rightarrow \overline{r}$ in the equations for the initial data, where $\overline{r}$ is rigged to satisfy $\overline{r}(0)=r_0$, $\overline{r}(r_t)=r_t$, and $\overline{r}=r$ for $r > r_t$. In addition, we require $\overline{r}$ to have a specified number of continuous derivatives (typically $8$), such that the turduckened and original data match to the specified degree of smoothness where they are joined. A polynomial $\overline{r}$ with these properties is constructed by solving a system of linear equations for the polynomial coefficients. 

\begin{figure}
 \begin{center}
 \includegraphics[width=3.4in]{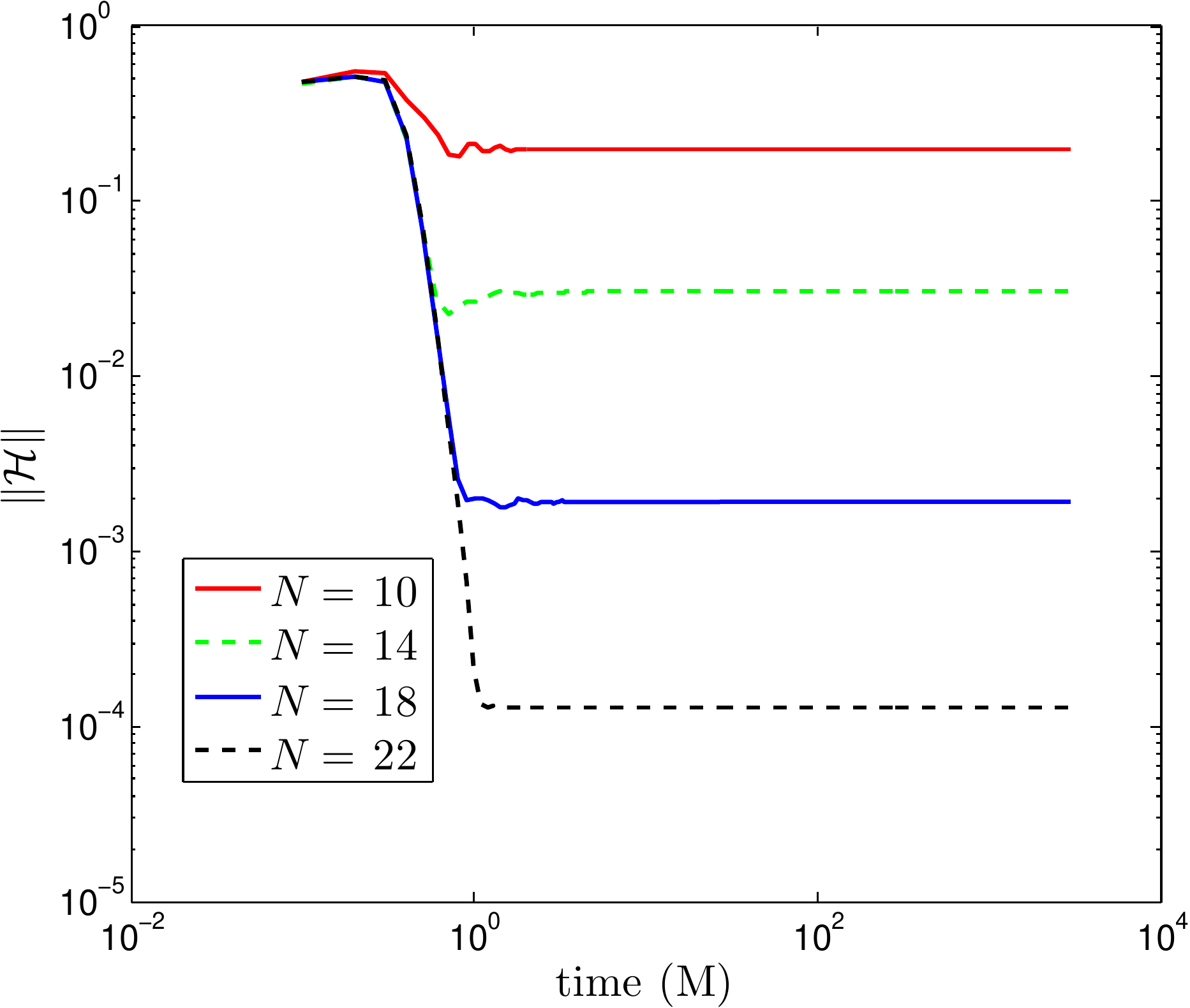}
 \end{center}
 \caption{Evolutions of a black hole in spherical symmetry using the standard second order in space BSSN formulation and a dG scheme with turduckening, see Section \ref{sec:turducken} for details. The norm used is the $L_2$ one. }
 \label{fig:FOBSSN_2ndorderstability}
 \end{figure}

In effect, our prescription stretches the physically correct (non-singular) data for the region $[r_0, r_t]$ over the turduckened region $[0, r_t]$.  This choice of initial data will naturally be constraint violating. As the constraint system's wavespeeds (see Section~\ref{sec:hyperbolicity} and Ref.~\cite{Brown:2008sb}) are not superluminal, these violations remain ``trapped" inside the horizon for all future times. Furthermore, for the second order BSSN system, Ref.~\cite{Brown:2008sb} found that the region of constraint violation quickly shrinks relative to the numerical grid. We experimented with turduckening the second order dG scheme described in \cite{Field:2010mn}, and found that the region of constraint violation quickly shrinks with this scheme as well. These simulations used a grid with a larger outer boundary and staggered domain sizes: 1 subdomain $[0, r_t=.4]M$ comprising the turduckened region with $r_0=.1M$, 3 subdomains in $[.4, 1.5]M$, 6 subdomains in $[1.5, 10]M$, and 12 subdomains in $[10, 100]M$. Otherwise the same numerical settings as in ~\ref{sec:ResultsDG}, although here we use frozen outer boundary conditions on all fields and gauge source terms chosen so that Eqs.~(\ref{eq:bm-slicing}, \ref{gammadrivershifteqns}) are initially time-independent. Fig.~\ref{fig:FOBSSN_2ndorderstability} shows the scheme is stable and whenever $t > M$ converges towards a time-independent solution exponentially with $N$.

Furthermore, the technique is observed to be robust for a variety of numerical parameter choices and domain decompositions. Results from our second order BSSN dG code suggest the turduckening technique to be applicable beyond finite difference schemes. Nevertheless, we have been largely unsuccessful at achieving robust stability turducken tests for FOBSSN. A typical evolution lasts on the order of tens to hundreds of $M$, although some low resolution runs can last into the thousands of $M$ before crashing.

\begin{figure*}
\begin{center}
\includegraphics[width=1.5\columnwidth,height=2in]{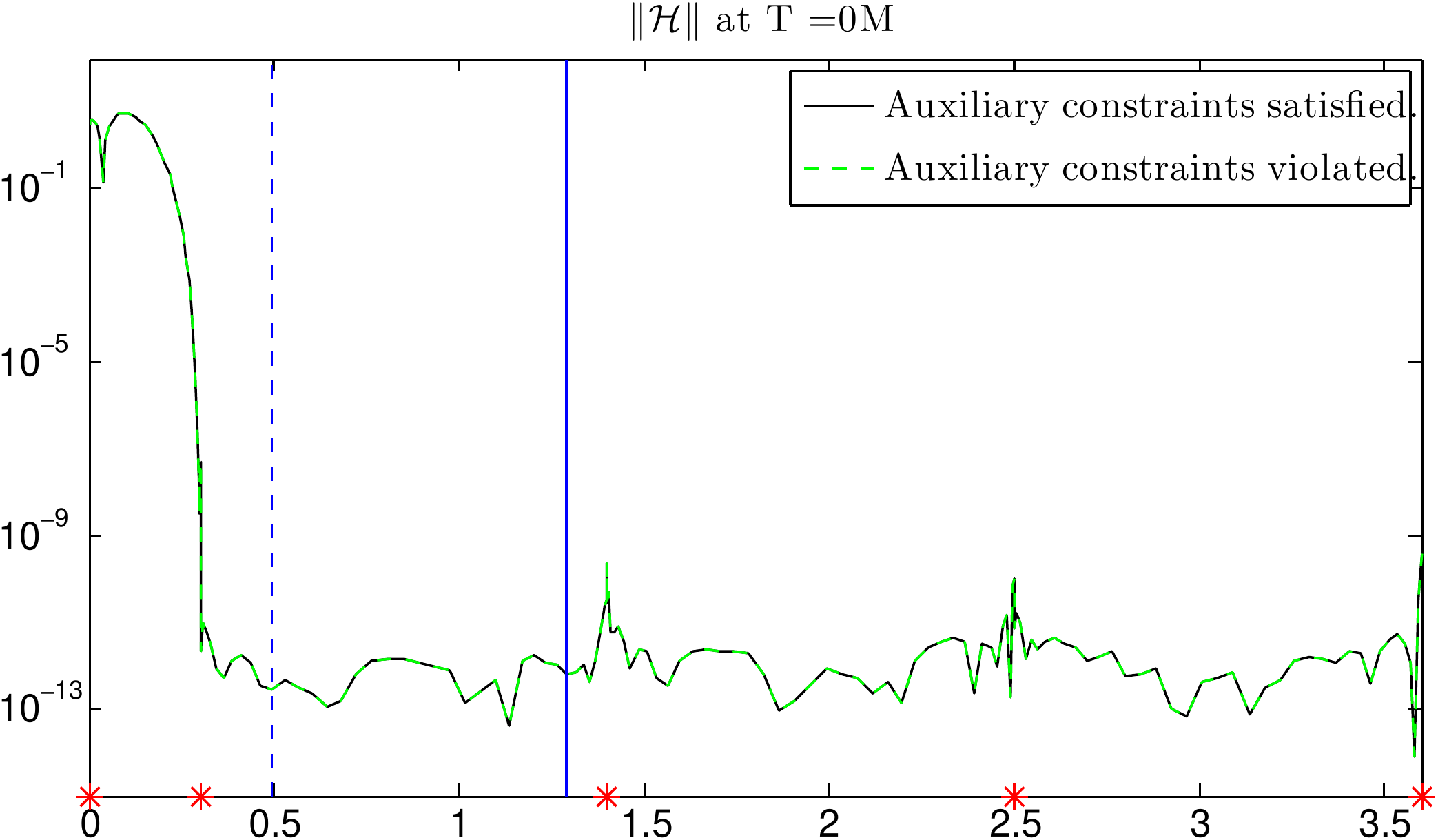} \\
\vspace{10pt}
\includegraphics[width=1.5\columnwidth,height=2in]{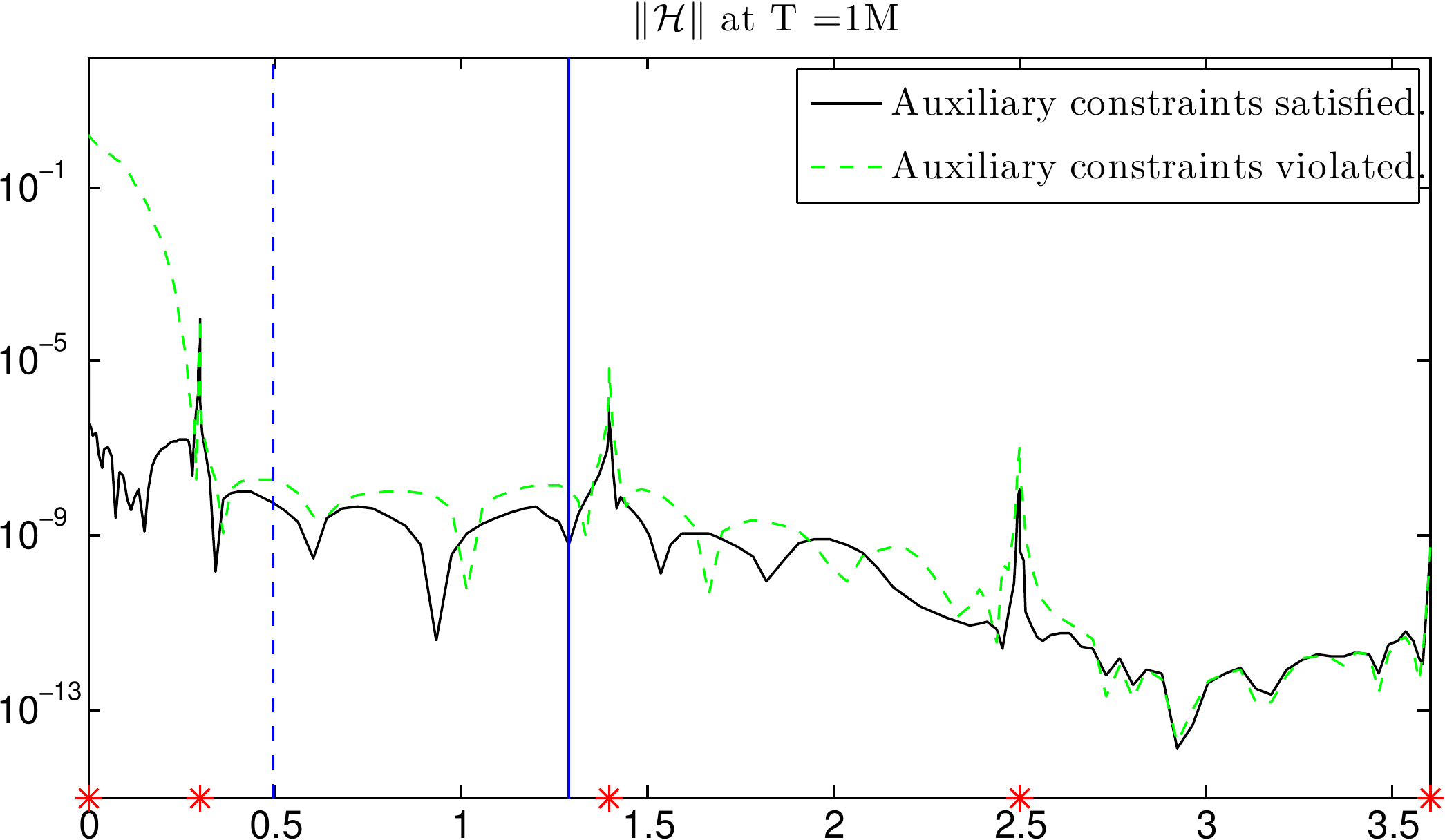} \\
r($M$)
\end{center}
\caption{Snapshots of the Hamiltonian constraint for turduckened initial data which satisfies (black line) and violates (dashed green line) the auxiliary constraint conditions~(\ref{Dconstraint})-(\ref{Aconstraint}) which arise in the first order reduction. Notice that only in the first case the constraint violating region ``shrinks'' in time, see Section \ref{sec:turducken} for more details. 
The solid vertical line marks the location of the event horizon, and the dashed one the location of the outermost radius with purely outgoing modes (where excision could be performed). The large red asterisks on the horizontal axis mark the location of each subdomain boundary.}
\label{fig:TurduckenConstraintPropogation}
\end{figure*}

The presence of extra auxiliary constraints (\ref{Dconstraint})-(\ref{Aconstraint}) presents a genuine difference between turduckening a first and second order BSSN system. In the first order system we have two distinct choices for calculating the auxiliary variables in the turduckened region of the initial data: calculating the analytic derivatives of the fields at the turduckened grid points or applying the numerical derivative operator to the turduckened fields. In the first case the auxiliary constraints are violated since the auxiliary fields correspond to derivatives of the non-turduckened fields. In the second case the auxiliary constraints are satisfied but the turduckened initial data no longer represents the physically correct data for $[r_0, r_t]$ stretched over region $[0, r_t]$. 
We experimented with both choices and found that the region of constraint violation is not guaranteed to shrink when using the first choice in which the auxiliary constraints are violated. Fig.~\ref{fig:TurduckenConstraintPropogation} documents a typical comparison with turduckening parameters $r_t=.3M$ and $r_0=.1M$. We note that our observations are influenced, for example, by the source terms $S_\alpha$, $S_\beta^r$, and $S_B^r$. 

\section{Comments}\label{sec:Conclusions}

The goal of this paper has been  a first step towards combining the robustness and simplicity of evolutions of the BSSN formulation of Einstein's equations, most notably being able to avoid the complications of excision, with very high accuracy numerical schemes -- those being a multi-domain pseudo spectral collocation method and a discontinuous Galerkin method. Furthermore, any of these approaches would allow, due to their memory efficiency and the speedup of GPUs (Graphics Processing Units), to run binary black hole simulations on a single GPU, thereby avoiding the bottleneck of PCIe communication between CPUs; see for example \cite{Bruegmann:2011zj}.  

For this purpose we derived and analyzed the hyperbolicity, characteristic variables and constraint propagation of a fully first order BSSN formulation of the Einstein equations with optional constraint damping terms, FOBSSN. Unfortunately, we have not been able to derive a symmetric hyperbolic formulation, but only a strongly hyperbolic one. It is known that in more than one spatial dimension, strong hyperbolicity, even with maximal dissipative boundary conditions, does not guarantee well posedness of the initial-boundary value problem~\cite{Calabrese:2003as}.  Yet, in our numerical experiments we have been able to carry out binary black hole simulations using our FOBSSN system, finite differences and adaptive mesh refinement, without any need for fine-tuning and no obvious signs of 
time instability (convergent errors that grow in time) or numerical instability (errors that get larger at higher resolutions at any fixed time).
Most notably, the presence of the extra constraints in FOBSSN seem to cause no problems. 

Next natural steps would be three-dimensional discontinuous Galerkin evolutions of FOBSSN using, for example, Hedge \cite{hedgeweb}, and implementation of more sophisticated boundary conditions, such as those of Ref.~\cite{Nunez:2009wn}. 

\begin{acknowledgments}
  We thank C\'eline Catto\"en, Lawrence Kidder, Stephen Lau, Rob Owen,
  Nicholas Taylor, and Saul Teukolsky for helpful discussions. This
  research has been supported by a Karen T. Romer Undergraduate
  Teaching and Research Award at Brown, a grant from the Sherman
  Fairchild Foundation to Cornell, the Center for Computation \&
  Technology at Louisiana State University, and the Joint
  Space-Science Institute at the University of Maryland. We
  acknowledge funding via NSF grants PHY-0652952, DMS-0553677,
  PHY-0652929, and NASA grant NNX09AF96G to Cornell, NSF awards
  PHY-0701566, OCI-0721915, OCI-0725070, PHY-0904015, OCI-0905046,
  OCI-0941653 to Louisiana State University, NSF awards PHY-0801213
  and PHY-1005632 to the University of Maryland, NSF award 
  PHY-0758116 to North Carolina State University, 
  and Grants CONACyT 61173 and CIC 4.19 to Universidad Michoacana. Calculations for this
  work were carried out at Compute Canada under allocation xck-093-ab,
  at LONI under allocation loni\_numrel06, at NERSC under allocation
  m152, and on the TeraGrid/XSEDE under allocations TG-MCA02N014 and
  TG-PHY090080.
\end{acknowledgments}


\appendix

\section{The covariant BSSN system} \label{app:BSSN}
In this appendix we sketch the derivation of the covariant BSSN system, Eqs.~(\ref{eq:evolBSSN}). The derivation follows 
from the analysis in Ref.~\cite{Brown:2009dd} by setting the determinant of the conformal metric to $\overline\gamma$, and choosing the 
trace of $\tilde A_{ij}$ to vanish. For simplicity we assume that the 
fiducial fields $\overline\gamma$ and $\overline\Gamma^i{}_{jk}$ are constructed from a time--independent metric $\overline\gamma_{ij}$. 
Unlike in the main body of the paper, here we do not assume that the fiducial metric is flat. 

Begin with the evolution equations for the physical spatial metric and extrinsic curvature,
\begin{subequations}\label{admeqns}
\begin{eqnarray}
        \partial_\perp \gamma_{ij} & = & -2\alpha K_{ij} \ ,\\
        \partial_\perp K_{ij} & = & \alpha\left[ R_{ij} - 2K_{ik} K^k_j + K K_{ij} \right] \nonumber \\
                & &  - D_i D_j \alpha  \ ,
\end{eqnarray}
\end{subequations}
where $R_{ij}$ and $D_i$ are the Ricci tensor and covariant derivative 
for $\gamma_{ij}$. Now let $\partial_\perp \equiv \partial_t - {\cal L}_\beta$ act on the BSSN variables $\phi$ and $\tilde\gamma_{ij}$, 
which are defined in 
Eqs.~(\ref{BSSNvariabledef}). The right--hand 
sides of these equations are written in terms of BSSN variables by inverting the definitions (\ref{BSSNvariabledef}):
\begin{subequations}\label{inverseBSSNdefs}
\begin{eqnarray}
        \gamma_{ij} & = & e^{4\phi}\tilde\gamma_{ij} \ ,\\
        K_{ij} & = & e^{4\phi}\tilde A_{ij} + \frac{1}{3} \gamma_{ij} K \ .
\end{eqnarray}
\end{subequations}
The results are identical 
to Eqs.~(\ref{eq:evolBSSN}a) and (\ref{eq:evolBSSN}b), respectively. 

The derivation of the equation of motion (\ref{eq:evolBSSN}d) for $\tilde A_{ij}$ follows the same pattern; apply $\partial_\perp$ to 
$\tilde A_{ij}$ in Eq.~(\ref{BSSNvariabledef}d), use the evolution Eqs.~(\ref{admeqns}), then replace the physical metric 
and extrinsic curvature with the BSSN variables through Eqs.~(\ref{inverseBSSNdefs}). 
In this case we 
must also write the physical Ricci tensor $R_{ij}$ in terms of conformal variables. Insert the 
relation 
\begin{equation}
        \Gamma^i{}_{jk} = \tilde\Gamma^i{}_{jk} + 2(\delta_j^i\tilde D_k\phi + \delta_k^i\tilde D_j\phi - \tilde\gamma_{jk}\tilde D^i\phi )
\end{equation}
for the Christoffel symbols into the definition of the Ricci tensor. This yields the splitting (\ref{Ricciconformalsplit}) between the 
conformal Ricci tensor $\tilde R_{ij}$ and the terms (\ref{Ricciphiterms}) that depend on the conformal factor $\phi$. 

The derivation of the identity (\ref{conformalRicci}) used
for the conformal Ricci tensor is somewhat tedious. Beginning with the definition 
\begin{equation}
        \tilde R_{ij} = \partial_k \tilde\Gamma^k{}_{ij} - \partial_i \tilde\Gamma^k{}_{jk} 
                + \tilde\Gamma^k{}_{ij}\tilde\Gamma^l{}_{kl} - \tilde\Gamma^k{}_{il}\tilde\Gamma^l{}_{jk}  
\end{equation}
it is staightforward to show that the difference between the conformal and fiducial Ricci tensors is 
\begin{eqnarray}\label{Riccidifference}
        \tilde R_{ij} - \overline R_{ij} & = & \overline D_k\Delta\tilde\Gamma^k{}_{ij} - \overline D_i\Delta\tilde\Gamma^k{}_{jk} 
                \nonumber \\ & & 
                + \Delta\tilde\Gamma^k{}_{ij}\Delta\tilde\Gamma^l{}_{kl} - \Delta\tilde\Gamma^k{}_{il}\Delta\tilde\Gamma^l{}_{jk} \ .
\end{eqnarray}
One can also show that 
\begin{equation}\label{deltatildegammaeqn}
        \Delta\tilde\Gamma^i{}_{jk} = \frac{1}{2} \tilde\gamma^{il} \left( \overline D_j\tilde\gamma_{kl} 
                + \overline D_k \tilde\gamma_{jl} - \overline D_l\tilde\gamma_{jk} \right) \ ,
\end{equation}
and derive the useful relations $\overline D_k \tilde\gamma_{ij} = 2\Delta\tilde\Gamma_{(ij)k}$ and 
$\overline D_k \tilde\gamma^{ij} = -2\Delta\tilde\Gamma^{(ij)}{}_k$. 
With these results, the first two terms in the difference (\ref{Riccidifference}) become
\begin{widetext}
\begin{eqnarray}
        \overline D_k\Delta\tilde\Gamma^k{}_{ij} - \overline D_i\Delta\tilde\Gamma^k{}_{jk} 
        & = & -\frac{1}{2} \tilde\gamma^{kl}\overline D_k\overline D_l \tilde\gamma_{ij} + \overline D_{(i}(\Delta\tilde\Gamma_{j)k}{}^k) 
        - (\overline D_{(i}\tilde\gamma^{kl})(\overline D_k\tilde\gamma_{j)l}) - \frac{1}{2} (\overline D_k\tilde\gamma^{kl})(\overline D_l\tilde\gamma_{ij}) \nonumber\\
        & & + (\overline D_k\tilde\gamma^{kl})(\overline D_{(i}\tilde\gamma_{j)l}) - \overline R_{ij} - \tilde\gamma^{kl}\tilde\gamma_{m(i}\overline R_{j)kl}{}^m \ .
\end{eqnarray}
With the definition $\Delta\tilde\Gamma^i \equiv \tilde\gamma^{jk}\Delta\tilde\Gamma^i{}_{jk}$, 
the conformal Ricci tensor from Eq.~(\ref{Riccidifference}) becomes 
\begin{eqnarray}
        \tilde R_{ij} & = & -\frac{1}{2} \tilde\gamma^{kl} \overline D_k\overline D_l \tilde\gamma_{ij} 
                + \tilde\gamma_{k(i}\overline D_{j)}\Delta\tilde\Gamma^k - \tilde\gamma^{kl}\tilde\gamma_{m(i}\overline R_{j)kl}{}^m 
                + \tilde\gamma^{kl}\Delta\tilde\Gamma^m{}_{kl} \Delta\tilde\Gamma_{(ij)m} \nonumber\\ & & 
                + \tilde\gamma^{kl}(2\Delta\tilde\Gamma^m{}_{k(i}\Delta\tilde\Gamma_{j)ml} + \Delta\tilde\Gamma^m{}_{ik}\Delta\tilde\Gamma_{mjl}) \ .
\end{eqnarray}
\end{widetext}
If the fiducial metric is flat, as assumed in the main body of the paper, then the fiducial Riemann tensor term on the right--hand 
side vanishes. The result (\ref{conformalRicci}) is obtained by replacing $\Delta\tilde\Gamma^i$ with the new variable
$\tilde\Lambda^i$ and dropping the fiducial Riemann tensor. 

To obtain the equation of motion (\ref{eq:evolBSSN}c) for $K$, we first let $\partial_\perp$ act on $K \equiv \gamma^{ij} K_{ij}$, using 
the results (\ref{admeqns}). The right--hand side is simplified by adding $-\alpha {\cal H}$, where 
${\cal H} = K^2 - K_{ij}K^{ij} + R$ is the Hamiltonian constraint. Equation (\ref{eq:evolBSSN}c) then follows 
after using the inverse relations (\ref{inverseBSSNdefs}) to write the result in terms of BSSN variables.

The conformal connection vector is defined in Eq.~(\ref{eq:Gamma}). To derive the evolution equation (\ref{eq:evolBSSN}e) 
for $\tilde\Lambda^i$, 
we first let the operator $\partial_\perp$ act on $\Delta\tilde\Gamma^i\equiv \tilde\gamma^{jk}\Delta\tilde\Gamma^i{}_{jk}$ 
with $\Delta\tilde\Gamma^i{}_{jk}$ expressed as in 
Eq.~(\ref{deltatildegammaeqn}). This generates several terms of the form $\partial_\perp (\overline D_i\tilde\gamma_{jk})$. Using the 
fact that Lie derivatives and partial derivatives commute \cite{Brown:2005aq} one can write these as 
\begin{equation}
        \partial_\perp (\overline D_i \tilde\gamma_{jk}) = \overline D_i(\partial_\perp \tilde\gamma_{jk}) 
                + 2 ({\cal L}_\beta\overline\Gamma^l{}_{i(j}) \tilde\gamma_{k)l} \ .
\end{equation}
Now use the identity \cite{Schouten_book}
\begin{equation}
        {\cal L}_\beta \overline\Gamma^i{}_{jk} = \overline D_j \overline D_k \beta^i - \overline R^i{}_{jkl}\beta^l \ ,
\end{equation}
which is straightforward to verify. The result of this calculation for $\partial_\perp\Delta\tilde\Lambda^i$ 
is an expression in which the operator $\partial_\perp$ 
acts only on the conformal metric $\tilde\gamma_{ij}$. Using the equation of motion (\ref{eq:evolBSSN}b), we find
\begin{widetext}
\begin{equation}
        \partial_\perp(\Delta\tilde\Gamma^i) = \tilde\gamma^{jk}\overline D_j\overline D_k\beta^i - \tilde\gamma^{jk}\overline R^i{}_{jkl}\beta^l
                - \frac{2}{\sqrt{\overline\gamma}} \overline D_j\left(\alpha \sqrt{\overline\gamma} \tilde A^{ij}\right) 
                + \frac{2}{3} \tilde\gamma^{jk}\Delta\tilde\Gamma^i{}_{jk} \overline D_l\beta^l 
                + \frac{1}{3}\tilde D^i ( \overline D_k\beta^k) \ .
\end{equation}
Next we add the term $2\alpha \tilde{\cal M}^i$, which is proportional to the momentum constraint (\ref{secondorderconstraints}b), to obtain 
\begin{eqnarray}
        \partial_\perp(\Delta\tilde\Gamma^i) & = & \tilde\gamma^{jk}\overline D_j\overline D_k\beta^i - \tilde\gamma^{jk}\overline R^i{}_{jkl}\beta^l
                + \frac{2}{3}\tilde\gamma^{jk}\Delta\tilde\Gamma^i{}_{jk} \overline D_\ell\beta^\ell 
		+ \frac{1}{3}\tilde D^i(\overline D_k\beta^k) \nonumber\\ & & 
		- 2\tilde A^{ik} \partial_k\alpha
		+ 2\alpha \tilde A^{k\ell}\Delta\tilde\Gamma^i{}_{k\ell}  
		+ 12\alpha \tilde A^{ik}\partial_k\phi
		- \frac{4}{3} \alpha \tilde D^i  K \ .
\end{eqnarray}
\end{widetext}
Now use the definition (\ref{eq:Gamma}) to replace $\partial_\perp\Delta\tilde\Gamma^i$ with $\partial_\perp\tilde\Lambda^i$. 
The result, assuming the fiducial metric is flat so that $\overline R^i{}_{jkl}$ vanishes, is the equation of motion 
(\ref{eq:evolBSSN}e).

\vspace{8ex}

\section{Fundamental variables in terms of characteristic variables}
The first--order BSSN variables in the gauge block can be obtained from the
characteristic variables using the formulas
\begin{widetext}
\begin{subequations}
\begin{eqnarray}
B_A &=& \frac{1}{2}\left( G_A^{(+)} + G_A^{(-)} \right)
 - \frac{\kappa^\beta}{\mathring\alpha^2 \mathring G}\beta_A \ ,\\
\beta_{nA} &=& \beta_{An} + \frac{\mathring\alpha}{2}\sqrt{\frac{\mathring G}{\mathring H}}
 \left( G_A^{(+)} - G_A^{(-)} \right) \ ,\\
K &=& \frac{1}{2}\left( G^{(\alpha,+)} + G^{(\alpha,-)} \right)
  + \frac{\kappa^\alpha}{\mathring\alpha^2 \mathring f}\, \alpha \ ,\\
\alpha_n &=& -\frac{\mathring\alpha\sqrt{\mathring f}}{2}
  \left( G^{(\alpha,+)} - G^{(\alpha,-)} \right) \ ,\\
B_n &=& \frac{1}{2}\left( G^{(\beta,+)} + G^{(\beta,-)} \right)
  - \frac{\kappa^\beta}{\mathring\alpha^2 \mathring G}\, \beta_n
  - \frac{4\mathring H}{3(\lambda^2-\mathring f)}\frac{1}{\mathring\alpha} \alpha_n \ ,\\
\beta_{nn} &=& -\beta_{AB}\delta^{AB}
+ \frac{\mathring\alpha \mathring G}{\lambda}\left[ 
  \frac{1}{2}\left( G^{(\beta,+)} - G^{(\beta,-)} \right)
 + \frac{4\mathring H}{3(\lambda^2-\mathring f)}\left( 
 \lambda K - \frac{\kappa^\alpha}{\mathring\alpha^2\lambda}\, \alpha \right) \right] \ .
\end{eqnarray}
\end{subequations}
\end{widetext}
We then apply the relations 
\begin{subequations}
\begin{eqnarray}
\alpha_i &=& n_i\alpha_n + \mathring\gamma_i{}^A\alpha_A \ , \\
\beta_{ij} &=& \beta_{nn} n_i n_j + n_i\mathring\gamma_j{}^A\beta_{nA} \nonumber \\&&
 + \mathring\gamma_i{}^A n_j\beta_{An} + \mathring\gamma_i{}^A\mathring\gamma_j{}^B\beta_{AB} \ ,
\end{eqnarray}
\end{subequations}
with similar expressions for $\beta_i$ and $B_i$.

For the non--gauge block, the inverse transformation is 
\begin{widetext}
\begin{subequations}
\begin{eqnarray}
\tilde{A}_{AB}^{tf} &=& \frac{1}{2}
  \left( {V}_{AB}^{(+)} + {V}_{AB}^{(-)} \right)
 + \frac{\kappa^\gamma}{2\mathring\alpha}\,\tilde{\gamma}_{AB}^{tf}
 + \frac{1}{\mathring\alpha}\, e^{-4\mathring\phi}{\beta}_{(AB)}^{tf} \ ,\\
\tilde{\gamma}_{nAB}^{tf} &=& -\left( {V}_{AB}^{(+)} - {V}_{AB}^{(-)} \right) \ , \\
\tilde{\Lambda}_A &=& \frac{1}{\mathring H}\left( Z_A + B_A \right) \ ,\\
\tilde{A}_{nB} &=& \frac{1}{2}\left( V_{nB}^{(+)} + V_{nB}^{(-)} \right)
 + \frac{\kappa^\gamma}{2\mathring\alpha}\,\tilde{\gamma}_{nB}
 + \frac{1}{\mathring\alpha}\, e^{-4\mathring\phi}\beta_{Bn} \ ,\\
\tilde{\gamma}_{nnB} &=& -\left( V_{nB}^{(+)} - V_{nB}^{(-)} \right)
 + e^{-4\mathring\phi}\left( \tilde{\Lambda}_B - 2\phi_B 
   - \frac{1}{\mathring\alpha}\alpha_B \right) \ ,\\
\tilde{\Lambda}_n &=& \frac{1}{\mathring H}\left( Z_n + B_n \right) \ ,\\
\phi_n &=& Z_0 + \frac{1}{8}\,\tilde{\Lambda}_n \ ,\\
\tilde{A}_{nn} &=& \frac{1}{2}\left( V_{nn}^{(+)} + V_{nn}^{(-)} \right)
 + \frac{\kappa^\gamma}{2\mathring\alpha}\,\tilde{\gamma}_{nn}
 + \frac{2}{3}\, e^{-4\mathring\phi} K
 - \frac{1}{\mathring\alpha}\, e^{-4\mathring\phi}\beta_{AB}\delta^{AB} \ ,\\
\tilde{\gamma}_{nnn} &=& -\left( V_{nn}^{(+)} - V_{nn}^{(-)} \right)
 + \frac{4}{3}\, 
   e^{-4\mathring\phi}\left( \tilde{\Lambda}_n - 2 \phi_n \right) \ .
\end{eqnarray}
\end{subequations}
The tensor components $\tilde A_{ij}$ and $\tilde \gamma_{kij}$ are then reconstructed as 
\begin{subequations}
\begin{eqnarray}
\tilde{A}_{ij} &=& 
 \tilde{A}_{nn}\left( \frac{3}{2}\, n_i n_j - \frac{1}{2}\mathring\gamma_{ij} \right)
 + n_i\mathring\gamma_j{}^B\tilde{A}_{nB} + n_j\mathring\gamma_i{}^B\tilde{A}_{nB} 
 + \mathring\gamma_i{}^A\mathring\gamma_j{}^B\hat{A}_{AB}  \ ,\\
\tilde{\gamma}_{kij} &=& \tilde{\gamma}_{nnn}
n_k\left( \frac{3}{2}\, n_i n_j - \frac{1}{2}\mathring\gamma_{ij} \right)
 + n_k n_i\mathring\gamma_j{}^B\tilde{\gamma}_{nnB} + n_k n_j\mathring\gamma_i{}^B\tilde{\gamma}_{nnB} 
 + n_k\mathring\gamma_i{}^A\mathring\gamma_j{}^B\tilde{\gamma}_{nAB}^{tf} + \mathring\gamma_k{}^A\tilde{\gamma}_{Aij} \ .
\end{eqnarray}
\end{subequations}
\end{widetext}


\bibliographystyle{physrev}
\bibliography{references,einsteintoolkit}

\end{document}